\documentclass[aps,preprint]{revtex4}%
\usepackage{amsfonts}
\usepackage{amsmath}
\usepackage{amssymb}
\usepackage{epsfig}
\usepackage{graphicx}%
\begin{document}
\title{Optimized Baxter Model of Protein Solutions: Electrostatics versus Adhesion}
\author{Peter Prinsen and Theo Odijk*,}
\affiliation{Complex Fluids Theory, Faculty of Applied Sciences, Delft University of
Technology, Delft, the Netherlands}

\begin{abstract}
A theory is set up of spherical proteins interacting by screened
electrostatics and constant adhesion, in which the effective adhesion
parameter is optimized by a variational principle for the free energy. An
analytical approach to the second virial coefficient is first outlined by
balancing the repulsive electrostatics against part of the bare adhesion. A
theory similar in spirit is developed at nonzero concentrations by assuming an
appropriate Baxter model as the reference state. The first-order term in a
functional expansion of the free energy is set equal to zero which determines
the effective adhesion as a function of salt and protein concentrations. The
resulting theory is shown to have fairly good predictive power for the
ionic-strength dependence of both the second virial coefficient and the
osmotic pressure or compressibility of lysozyme up to about $0.2$ volume
fraction.\vspace{3.6in}

Address for correspondence: T. Odijk, P.O. Box 11036 2301 EA Leiden, the Netherlands

\end{abstract}
\maketitle

\section*{I. INTRODUCTION}

It has been intimated that the solution properties of globular proteins may
bear relation with their crystallization properties \cite{GE1,GE2}. Since the
characterization of proteins commands ever more attention, such a contention
is of considerable interest so much work has been carried out on this topic
recently \cite{HAG,ROS,HA1,HA2,HA3,NEA}.

The difficulty of setting up a predictive theory of protein suspensions based
on what is known about the interaction between two proteins, has been
acknowledged for some time \cite{VIL}. Best fitting of the osmotic pressure
of, for instance, bovine serum albumin up to $100$ $%
\operatorname{g}%
/%
\operatorname{l}%
$, leads to effective excluded volumes whose behavior as a function of salt is
enigmatic \cite{MIN}.

In recent years, there has been a tendency to forget about all detail of the
protein interaction altogether---both attractive and repulsive---and simply
introduce a single adhesion parameter \cite{MIN,FIN,LOM,PIA2,ROS2}. Despite
the electrostatic repulsion which is substantial, the data are often merely
rationalized in terms of the bare protein diameter within the context of an
adhesive sphere model and such an approach seems to have merit
\cite{MIN,FIN,LOM,PIA2,ROS2}. This empiricism has prompted us to develop a
theory of screened charged protein spheres that have a constant stickiness,
where the electrostatic interaction is compensated, in part, by the adhesive
forces. Thus, we argue that, effectively, the spheres are asigned a hard
diameter identical to the actual diameter provided the remnant adhesive
interaction now depends on the electrolyte and protein concentrations in a
manner to be determined variationally. First, we analyze the second virial
coefficient as such, for this will point toward a way of dealing with the
osmotic pressure at nonzero concentrations. We focus on experiments with
lysozyme, a protein which is reasonably spherical and has been well studied
for a long time \cite{SOP}.

\section*{II. SECOND VIRIAL COEFFICIENT}

\subsection*{A. Theory}

\subsection*{\textit{1. Second virial coefficient}}

The second virial coefficient $B_{2}$ describes the first order correction to
Van 't Hoff's law%

\begin{equation}
\frac{\Pi}{\rho k_{B}T}=1+B_{2}\rho+O(\rho^{2}).
\end{equation}
Here, $\Pi$ is the osmotic pressure of the solution, $\rho$ is the particle
number density, $k_{B}$ is Boltzmann's constant and $T$ is the temperature.
From statistical mechanics we know that, given the potential of mean force
$U\left(  \mathbf{r}\right)  $ between two spherical particles whose centers
of mass are separated by the vector $\mathbf{r}$, one can calculate $B_{2}$ from%

\begin{equation}
B_{2}=-\frac{1}{2}\int_{V}d\mathbf{r\;}f(\mathbf{r}) \label{vir}%
\end{equation}
where $f(\mathbf{r})=e^{-U\left(  \mathbf{r}\right)  /k_{B}T}-1$ is the Mayer
function. In principle, the interaction $U(\mathbf{r})$ may be determined from
experimental data on the second virial coefficient by suitable Laplace
inversion. This has been done for atoms and spherically symmetric molecules
\cite{KEL, MAI}, for which the second virial coefficient has been measured
over a broad enough range of temperatures. One might think of formulating a
procedure similar in spirit and applicable to protein solutions, but with the
ionic strength as independent variable instead of the temperature. However, to
be able to determine an accurate approximation of the interaction, the
experimental data have to be known fairly accurately, which is not the case at
hand, as will become clear further on. We are therefore forced to adduce
presumptions about the interaction.

We assume the protein to be spherical with radius $a$, its charge being
distributed uniformly on its surface. For convenience, all distances will be
scaled by the radius $a$ of the sphere and all energies will be in units of
$k_{B}T$. Because monovalent ions (counterions and salt ions) are also present
in solution, there will be a screened Coulomb repulsion between the proteins,
here given by a far-field Debye-H\"{u}ckel potential. We compute the effective
charge $qZ_{eff}$ in the Poisson-Boltzmann approximation where $q$ is the
elementary charge. For now, we let the attraction between two proteins be of
short range, and we model it by a potential well of depth $U_{A}$ and width
$\delta\ll1$. The total interaction $U(x)$ between two proteins is of the form%

\begin{equation}
U(x)=\left\{
\begin{array}
[c]{ll}%
\infty & 0\leq x<2\\
U_{DH}(x)-U_{A} & 2\leq x<2+\delta\\
U_{DH}(x) & x\geq2+\delta
\end{array}
\right.  , \label{pot}%
\end{equation}%
\[
x\equiv\frac{r}{a},
\]
with Debye-H\"{u}ckel potential \cite{VER}%
\begin{equation}
U_{DH}(x)=2\xi\frac{e^{-\mu(x-2)}}{x}. \label{debye}%
\end{equation}
Here, $\xi\equiv\frac{Q}{2a}\left(  \frac{Z_{eff}}{1+\mu}\right)  ^{2}$,
$\kappa^{-1}$ is the Debye length defined by $\kappa^{2}=8\pi QI$, $I$ is the
ionic strength, $Q=q^{2}/\epsilon k_{B}T$ is the Bjerrum length, which equals
$0.71$ $%
\operatorname{nm}%
$ in water at $298$ $%
\operatorname{K}%
$, $\epsilon$ is the permittivity of water and $\mu\equiv\kappa a=3.28a\sqrt
{I}$ in water, if $a$ is given in $%
\operatorname{nm}%
$ and $I$ in $%
\operatorname{mol}%
/%
\operatorname{l}%
$. We suppose 1-1 electrolyte has been added in excess so $I$ is the salt concentration.

In order to evaluate $B_{2}$ analytically, we have found it expedient to split
up $B_{2}$ into several terms:%

\begin{equation}
B_{2}=B_{2}^{HS}\left(  1+\frac{3}{8}J\right)  , \label{vir2}%
\end{equation}
where $B_{2}^{HS}=16\pi a^{3}/3$ is the second virial coefficient if the
proteins were merely hard spheres and we introduce integrals%
\begin{equation}
J\equiv\int_{2}^{\infty}dx\;x^{2}\left(  1-e^{-U\left(  x\right)  }\right)
\equiv J_{1}-\left(  e^{U_{A}}-1\right)  J_{2}, \label{vir3}%
\end{equation}%
\begin{equation}
J_{1}\equiv\int_{2}^{\infty}dx\;x^{2}\left(  1-e^{-U_{DH}(x)}\right)  ,
\end{equation}%
\begin{equation}
J_{2}\equiv\int_{2}^{2+\delta}dx\;x^{2}e^{-U_{DH}(x)}.
\end{equation}
Here, $J_{1}$ is the value of $J$ in the absence of attraction and may be
simplified by Taylor expanding the Boltzmann factor in the integrand for small
values of $U_{DH}$ to second order. However, to increase the accuracy of the
expansion, we adjust the coefficient of the second order term so that the
approximation to the integrand coincides with its actual value at $x=2$, i.e.,
we approximate $x\left(  1-e^{-U_{DH}(x)}\right)  \simeq2\xi e^{-\mu
(x-2)}-2\alpha\xi^{2}e^{-2\mu(x-2)}$, with $\alpha=\frac{e^{-\xi}-(1-\xi)}%
{\xi^{2}}$, resulting in%
\begin{equation}
J_{1}\simeq\frac{4\left(  \mu+\frac{1}{2}\right)  \xi}{\mu^{2}}\left(
1-\frac{\alpha}{2}\xi\right)  , \label{in1}%
\end{equation}
where we have neglected the small term $\alpha\xi^{2}/2\mu^{2}$. For instance,
in the case of lysozyme, the deviation of the approximation Eq. (\ref{in1})
from the exact result is smaller than about $3\%$ for $I\geq0.05$ M and
smaller than about $1\%$ for $I\geq0.2$ M. Since $\delta\ll1$, $J_{2}$ may be
simplified by using the trigonometrical approximation $\int_{2}^{2+\delta
}dx\;g(x)\simeq\frac{1}{2}\delta\left[  g(2)+g(2+\delta)\right]  $, which
leads to%
\begin{equation}
J_{2}\simeq2\delta\left[  e^{-\xi}+\left(  1+\frac{\delta}{2}\right)
^{2}e^{-\frac{\xi}{1+\delta/2}e^{-\mu\delta}}\right]  . \label{in2}%
\end{equation}
It is important to note that $\mu\delta$ may be greater than unity even if
$\delta\ll1$. Again, for lysozyme, this approximation deviates less than about
$3\%$ from the exact value for $I\geq0.2$ M and $\delta\leq0.5$ and less than
about $1\%$ for $I\geq0.2$ M and $\delta\leq0.15$.

\subsection*{\textit{2. Effective attractive well}}

We next present a discussion of $B_{2}$ in terms of equivalent interactions
and their Mayer functions even though the analysis of the previous section is
self-contained. Sections IIA2 and IIA3 may be viewed as preludes to the
formulation of the liquid-state theory developed in section III. At large
separations ($x>2+\delta$), the interaction between the particles is purely
repulsive, leading to a positive contribution to the second virial
coefficient. If, at a certain ionic strength, the second virial coefficient is
smaller than the hard-core value ($B_{2}<B_{2}^{HS}$), this positive
contribution is necessarily cancelled by only part of the negative
contribution of the attractive interaction at small separations, the part,
say, between $x=2+\epsilon_{0}$ and $x=2+\delta$, see Fig. 1. The remaining
potential which we will call an effective attractive well, then consists of a
hard-core repulsion plus a short-range attraction of range $\epsilon_{0}$. The
value of $\epsilon_{0}$ is determined by noting that the free energy of the
suspension must remain invariant, which, in the asymptotic limit of low
densities, leads to the identity%
\begin{equation}
B_{2,\epsilon_{0}}=B_{2},\label{idvir}%
\end{equation}
where $B_{2}\ $is the second virial coefficient of the previous section and
$B_{2,\epsilon_{0}}$ is the second virial coefficient pertaining to the
effective attractive well. Using Eq. (\ref{vir}), we rewrite Eq. (\ref{idvir})
as%
\begin{equation}
\int_{V}d^{3}\mathbf{r\;}\Delta f=0,\label{vrwma}%
\end{equation}
in terms of the difference in the respective Mayer functions%
\begin{equation}
\Delta f\equiv f-f_{\epsilon_{0}},
\end{equation}
where $f$ is the Mayer function of the original interaction and $f_{\epsilon
_{0}}$ is the Mayer function of the effective attractive well. In
dimensionless units, Eq. (\ref{vrwma}) is equivalent to the condition%

\begin{equation}
\int_{2+\delta}^{\infty}dx\;x^{2}\left(  1-e^{-U_{DH}(x)}\right)
=\int_{2+\epsilon_{0}}^{2+\delta}dx\;x^{2}\left(  e^{U_{A}}e^{-U_{DH}%
(x)}-1\right)  , \label{canc}%
\end{equation}
where, using the same approximation that led to Eq. (\ref{in1}), we write%
\begin{equation}
\int_{2+\delta}^{\infty}dx\;x^{2}\left(  1-e^{-U_{DH}(x)}\right)  \simeq
\frac{2\xi e^{-\mu\delta}}{\mu}\left(  1-\frac{\alpha}{2}\xi e^{-\mu\delta
}\right)  \left(  2+\delta+\frac{1}{\mu}\right)  \label{M0}%
\end{equation}
and, using $\int_{2+\epsilon_{0}}^{2+\delta}dx\;x^{2}\Delta f(x)\simeq
2(\delta-\epsilon_{0})\left[  \Delta f(2+\delta)+\Delta f(2+\epsilon
_{0})\right]  $,%
\begin{equation}
\int_{2+\epsilon_{0}}^{2+\delta}dx\;x^{2}\left(  e^{U_{A}}e^{-U_{DH}%
(x)}-1\right)  \simeq2(\delta-\epsilon_{0})\left[  -2+e^{U_{A}}\left(
e^{-\frac{\xi}{1+\delta/2}e^{-\mu\delta}}+e^{-\frac{\xi}{1+\epsilon_{0}%
/2}e^{-\mu\epsilon_{0}}}\right)  \right]  . \label{eps}%
\end{equation}
To leading order, we then find an explicit relation for $\varepsilon_{0}$%
\begin{equation}
\delta-\epsilon_{0}\simeq\frac{\xi e^{-\mu\delta}}{\mu e^{U_{A}}}e^{\xi
e^{-\mu\delta}}, \label{deteps}%
\end{equation}
which works well at high ionic strengths (i.e. at low values of $\xi$), e.g.
whenever $I\geq1$ M in the case of lysozyme at pH 4.5. A more accurate value
of $\delta-\epsilon_{0}$ is obtained by equating Eqs. (\ref{M0}) and
(\ref{eps}), and then iteratively updating the factor $(\delta-\epsilon_{0})$,
starting with the initial value $\epsilon_{0}=\delta$.

The second virial coefficient pertaining to the original potential $U(x)$ (Eq.
(\ref{pot})) is now rewritten as%
\begin{equation}
B_{2}=B_{2,\epsilon_{0}}=B_{2}^{HS}\left(  1+\frac{3}{8}\int_{2}%
^{2+\epsilon_{0}}dx\;x^{2}\left(  1-e^{U_{A}}e^{-U_{DH}\left(  x\right)
}\right)  \right)  .
\end{equation}
The depth $U_{A}-U_{DH}(x)$ does not vary strongly though, since
$\varepsilon_{0}\ll1$, so, to simplify things computationally, let us
approximate the interaction by a square well potential,%
\begin{equation}
U_{SW}(x)=\left\{
\begin{array}
[c]{ll}%
\infty & 0\leq x<2\\
-U_{S} & 2\leq x<2+\epsilon_{0}\\
0 & x\geq2+\epsilon_{0}%
\end{array}
\right.  .
\end{equation}
We choose $U_{S}$ in such a way that $B_{2}=B_{2}^{SW}$ or, equivalently,%
\begin{equation}
\int_{2}^{2+\epsilon_{0}}dx\;x^{2}\left(  e^{U_{S}}-e^{U_{A}}e^{-U_{DH}\left(
x\right)  }\right)  =0.
\end{equation}
To leading order in $\epsilon_{0}$, we have
\begin{equation}
\int_{2}^{2+\epsilon_{0}}dx\;x^{2}e^{U_{S}}\simeq4\epsilon_{0}e^{U_{S}},
\label{B2SW}%
\end{equation}
and, using the approximation $\int_{2}^{2+\epsilon_{0}}dx\;x^{2}%
g(x)\simeq2\epsilon_{0}\left[  g(2+\epsilon_{0})+g(2)\right]  $, we write%
\begin{equation}
\int_{2}^{2+\epsilon_{0}}dx\;x^{2}e^{U_{A}}e^{-U_{DH}\left(  x\right)  }%
\simeq2\epsilon_{0}e^{U_{A}}\left[  e^{-\xi}+e^{-\frac{\xi}{1+\epsilon_{0}%
/2}e^{-\mu\epsilon_{0}}}\right]  .
\end{equation}
The depth $U_{S}$ of the potential is then given by%
\begin{equation}
e^{U_{S}}\simeq\frac{1}{2}e^{U_{A}}\left(  e^{-\xi}+e^{-\frac{\xi}%
{1+\epsilon_{0}/2}e^{-\mu\epsilon_{0}}}\right)  \label{detus}%
\end{equation}
in terms of the original variables. Finally, we point out that the two
attractive wells that we have introduced are physically meaningful only if
$B_{2}<B_{2}^{HS}$.

\subsection*{\textit{3. Attractive well in the Baxter limit}}

We have shown that one may simplify the statistical thermodynamics of the
protein suspension at low densities considerably, by replacing the original
interaction, consisting of an electrostatic repulsion and a short-range
attraction, by a single attractive well of short range. The electrostatic
interaction may be substantial but it is compensated by part of the original
attractive well which is quite strong ($U_{A}>1$). Another useful interaction
expressing attractive forces of short range consists of a hard-sphere
repulsion and an attraction of infinite strength and zero range, namely the
adhesive hard sphere (AHS) potential of Baxter \cite{BAX}%

\begin{equation}
U_{AHS}(x)=\left\{
\begin{array}
[c]{ll}%
\infty & 0\leq x<2\\
\log\frac{12\tau\omega}{2+\omega} & 2\leq x\leq2+\omega\\
0 & x>2+\omega
\end{array}
\right.  , \label{AHS}%
\end{equation}
where $\tau$ is a constant and the limit $\omega\downarrow0$ has to be taken
after formal integrations. The second virial coefficient remains finite%
\begin{equation}
B_{2}^{AHS}=B_{2}^{HS}\left(  1-\frac{1}{4\tau}\right)  .
\end{equation}
Because much is known about the statistical mechanics of the Baxter model, one
often defines $\tau$ in terms of some $B_{2}$ and naively assumes there is a
one-to-one correspondence between the original and Baxter models. For
instance, in our case, $B_{2}^{AHS}=B_{2}=B_{2,\varepsilon_{0}}=B_{2}^{SW}$.
Since we have%
\begin{align}
B_{2}^{SW}  &  =B_{2}^{HS}\left(  1-\left(  e^{U_{S}}-1\right)  \left[
\left(  1+\frac{\epsilon_{0}}{2}\right)  ^{3}-1\right]  \right) \\
&  \simeq B_{2}^{HS}\left(  1-\frac{3}{2}\left(  e^{U_{S}}-1\right)
\epsilon_{0}\right)  ,\nonumber
\end{align}
we thus identify%
\begin{equation}
\frac{1}{\tau}\simeq6\epsilon_{0}\left(  e^{U_{S}}-1\right)  , \label{dettau}%
\end{equation}
where $U_{S}$ is given by Eq. (\ref{detus}). However, it is important to
realize that this procedure is legitimate at small densities only. At finite
concentrations, the optimal representation of the real suspension of proteins
by a Baxter model has to be derived and we will show in section III that the
simple-minded identification $B_{2}^{AHS}\equiv B_{2}$ no longer applies.

\subsection*{B. Application to lysozyme}

\subsection*{\textit{1. Experimental Data}}

Lysozyme is, by far, the best studied protein with regard to solution
properties. This is one of the reasons for using this protein to test theory,
another being its moderate aspect ratio of about 1.5 so that it may be fairly
well approximated by a sphere. Bovine Serum Albumin (BSA) has also been well
studied, but is considerably more anisometric with an aspect ratio of about
3.5. Numerous measurements of the second virial coefficient of lysozyme have
been published. In fact, there are quite a few sets of experiments pertinent
to our analysis \cite{ROS1, BEH, ROS2, BLO, BON, CUR1, CUR2, MUS, PIA1, VEL}.

It turns out that there is appreciable scatter in the data if we plot all
measurements of $B_{2}$ at a pH of about $4.5$ as a function of ionic strength
$I$ (NaCl + small amount of Na acetate; we have set the ionic strength arising
from the latter equal to $0.6\times$concentration \cite{VEL}) (see Fig. 2).
Several sets of data \cite{PIA1, BEH} appear to be way off the general curve
within any reasonable margin of error. An important criterion is how well the
$\theta$ point (i.e. when $B_{2}=0$) is established since then attractive
forces---which we would like to understand---are well balanced against
electrostatics---which we purportedly understand well. Experimentally
speaking, it ought to be possible to monitor $B_{2}$ accurately about the
$\theta$ point; large negative $B_{2}$ values at $I\gg I_{\theta}$ are more
difficult to determine because the proteins may start to aggregate or
nucleate, in principle. Various polynomial fits for all data close to the
$\theta$ point yield $I_{\theta}=0.20\pm0.01$ M. Hence, we have regarded data
sets \cite{PIA1, BEH} markedly disagreeing with this ionic strength as
anomalous so we have not taken them into consideration. Fig. 3 displays all
data we have taken into account. Clearly, the composite curve yields a fairly
reliable basis to test possible theories of the attractive force. On the other
hand, it is unclear at present how the scatter in data in Fig. 2 translates
into bounds for inferred attractive interactions.

\subsection*{\textit{2. Theory}}

\subsubsection*{\textit{2a. Electrostatics}}

Next, it is important to ascertain the actual and effective charges of
lysozyme under conditions relevant to the present work. Kuehner et al
\cite{KUE} performed hydrogen-ion titrations on hen-egg-white lysozyme in KCl
solutions. By interpolation, we obtain the actual charge $Z$ of the protein as
a function of the 1-1 electrolyte concentration $I$ (see tables I and II).
Experiments on $B_{2}$ are usually carried out with NaCl (and some Na acetate)
as the supporting monovalent electrolyte but we here assume KCl and NaCl
behave identically in an electrostatic sense.\ We solve the Poisson-Boltzmann
equation to get the effective charge $Z_{eff}$ in the Debye-H\"{u}ckel tail
(for more detail, see Appendix A). The dimensionless radius is set equal to
$\mu=3.28a\sqrt{I}=5.58\sqrt{I}$ and Eq.(\ref{cha}) is used to compute the
renormalized or effective charge. (Setting $a=1.7$ nm for lysozyme as in Refs.
\cite{ROS1} and \cite{CUR2}; the Bjerrum length $Q=0.71$ nm for H$_{2}$O at
room temperature). The other dimensionless parameter is given by
$\xi=0.209(\overline{Z}/(1+\mu))^{2}$, where $\overline{Z}=Z_{eff}-1$ (see below).

\subsubsection*{\textit{2b. Attractive well}}

We have assumed $U_{A}$ and $\delta$ to be independent of the ionic strength
$I$. It is possible to show that this does not contradict the data displayed
in Figs. 2 and 4. In Appendix B, we prove that if the interaction between the
proteins is given by Eq. (\ref{pot}), then $dB_{2}/d\mu<0$ and $d^{2}%
B_{2}/d\mu^{2}>0$, the last inequality being valid if $\xi<1$. We recall that
$\mu$ is proportional to $\sqrt{I}$ so that Figs. 3 and 4 indeed bear out
these inequalities after due rearrangement.

Next, we determine the optimal values of $U_{A}$ and $\delta$ yielding exact,
numerical $B_{2}(I)$ curves given by Eq. (\ref{vir2}) which are the best fits
to the data of Fig. 3. We require that $I_{\theta}=0.20\pm0.01$ is predicted
absolutely which fixes $U_{A}$, say, and $\delta$ is then determined by a
nonlinear minimization procedure. We thus obtain $U_{A}=1.70\pm0.25$ and
$\delta=0.468\mp0.097$ but we note that the quantity $\delta\exp U_{A}%
=2.56\pm0.10$ is much more narrowly bounded. Now, it can be argued that the
Debye-H\"{u}ckel potential with effective charge $Z_{eff}$ overestimates the
real potential in magnitude so we have repeated this numerical procedure with
a slightly lower effective charge, viz. $\overline{Z}=Z_{eff}-1$ (see tables I
and II). This yields the revised estimates $U_{A}=2.87\pm0.65$, $\delta
=0.167\mp0.086$ and $\delta\exp U_{A}=2.95\pm0.21$. The numerically computed
curves are displayed in Fig. 3. We therefore conclude that the variables
$U_{A}$ and $\delta$ as such are difficult to ascertain unambiguously, though
the variable $\delta\exp U_{A}$ is quite robust. This is also borne out if we
use our approximations, Eqs. (\ref{in1}) and (\ref{in2}), instead of the exact
numerical computations. There are again wide variations in $U_{A}$ and
$\delta$ but the quantity $\delta\exp U_{A}$ is strictly bounded: $\delta\exp
U_{A}=2.70\pm0.11$ (effective charge $=Z_{eff}$) and $\delta\exp U_{A}%
=3.02\pm0.21$ (effective charge $=Z_{eff}-1$).

We now argue why $\delta\exp U_{A}$ is indeed a relevant quantity, to a good
approximation. At the $\theta$ point we have $B_{2}=0$ so that $J_{\theta
}=-8/3$ from Eq. (\ref{vir2}). From tables I and II, we see that generally
$\mu\gg1$ and $\alpha\xi\ll1$; hence, we have $J_{1}\simeq4\xi/\mu$ and
$J_{2}\simeq4\delta\exp-\xi$ for often $\mu\delta>1$. This would lead to
$\delta\exp U_{A}\simeq4.4$. On the other hand, at very high $I$, $J_{1}$ and
$\xi$ tend to zero and, because $U_{A}\gg1$, the scaled virial coefficient
$B_{2}/B_{2}^{HS}$ reduces to $-\frac{3}{8}J_{2}\exp U_{A}\simeq-\frac{3}%
{2}\delta\exp U_{A}$ leading to $\delta\exp U_{A}\simeq3$ estimated from Fig.
3. Hence, the two estimates at the respective extremes are fairly consistent.
To summarize, we may propose a crude approximation to the second virial
coefficient which is a universal function of $\delta\exp U_{A}$%
\begin{equation}
\frac{B_{2}}{B_{2}^{HS}}\simeq1+\frac{3\xi}{2\mu}-\frac{3}{2}e^{-\xi}\delta
e^{U_{A}}. \label{benb2}%
\end{equation}
The third term on the right is exact in the limit $\delta\rightarrow0$,
whereas the absolute error in the second term is smaller than $0.25$ when
$I\geq0.1$ M. Using Eq. (\ref{benb2}) to fit the data leads to $\delta\exp
U_{A}=4.2$ when we use the effective charge $Z_{eff}$, whereas $\delta\exp
U_{A}=3.7$ when we use the lower effective charge $\overline{Z}$ (see Fig. 5).

In Fig. 3 we see that the curves at low values of $\delta$ fit the data at
high ionic strengths better. In the remainder of this article, we therefore
employ the values $\delta=0.079$ and $U_{A}=3.70$, corresponding to the
lowered effective charge $\overline{Z}$ and $I_{\theta}=0.21$ M. In Fig. 6 we
show a comparison between the experimental data at a pH of about 7.5 and the
theoretical curve computed numerically with the same parameters.

\subsubsection*{\textit{3. AHS potential}}

Values of $\epsilon_{0}$, $U_{S}$ and $\tau$ at several ionic strengths are
given in tables I and II. Fig. 7 displays the ionic-strength dependence of the
adhesion parameter $\tau$. Near the $\theta$ point, $\tau$ decreases quickly
with increasing $I$. At high ionic strength, $\tau$ approaches the limiting
value of $\left(  6\delta(e^{U_{A}}-1)\right)  ^{-1}$, which, upon the use of
our choice $\delta=0.079$ and $U_{A}=3.7$, is equal to $0.0535$. We note that
at pH $4.5$ and at ionic strengths $I=0.05$ M and $I=0.1$ M, the computed
values of $\epsilon_{0}$, $U_{S}$ and $\tau$ become nonsensical. In that case,
the attractive potential is simply not strong enough to compensate the
electrostatic repulsion completely so our analytical approach breaks down.
This can also be seen in Fig. 2, where we have $B_{2}>B_{2}^{HS}$ for these
two values of the ionic strength. The same effect occurs at pH $7.5$ when
$I=0.05$ M.

\section*{III. LIQUID\ STATE THEORY\ AT\ HIGHER DENSITIES}

\subsection*{A. Theory}

\subsection*{\textit{1. Density dependent attractive well in the Baxter
limit}}

In section II, we introduced the AHS potential as a convenient first
approximation to the interaction\ between proteins. We determined the adhesion
parameter $\tau$ by matching values of the second virial coefficient which is
methodologically correct only in the asymptotic limit of very low densities.
In this section we propose a new procedure of choosing $\tau$, which is valid
at higher densities but $\tau$ now depends on the protein density. We extend a
method originally proposed by Weeks, Chandler and Anderson \cite{AND} for
repulsive interactions. They variationally determined an effective hard sphere
diameter for a soft, repulsive potential of short-range, but we argue that
their scheme is more generally applicable as long as the full interaction
remains of short range, which is the case here.

We start by introducing a functional expansion\ of the excess Helmholtz free
energy $\Delta A$ in terms of the Mayer function of the interaction $U$%
\begin{align}
\rho^{-1}\mathcal{A}(\rho,T;\varphi_{s})  &  =\rho^{-1}\mathcal{A}%
(\rho,T;\varphi_{AHS})+\frac{\eta}{2}\frac{3}{4\pi}\int d\mathbf{x}%
\;B_{AHS}(x)+\label{WCA}\\
&  +\frac{\eta^{2}}{2}\left(  \frac{3}{4\pi}\right)  ^{2}\frac{a^{3}}{V}\int
d\mathbf{x}_{1}d\mathbf{x}_{2}d\mathbf{x}_{3}\;B_{AHS}(x_{12})B_{AHS}%
(x_{13})J_{AHS}^{(3)}(\mathbf{x}_{1},\mathbf{x}_{2},\mathbf{x}_{3}%
)+\ldots.\nonumber
\end{align}
Here, $V$ is the volume of the system, $\mathcal{A}=-\Delta A/V$, $\varphi
_{s}(x)=e^{-U(x)}$, $\varphi_{AHS}(x)=e^{-U_{AHS}(x)}$, $\eta=4\pi a^{3}%
\rho/3$ is the volume fraction of particles, $J_{AHS}^{(3)}(\mathbf{x}%
_{1},\mathbf{x}_{2},\mathbf{x}_{3})$ is a complicated function depending on
two and three particle correlation functions (see \cite{AND}) and
$\mathbf{x}_{12}=\mathbf{x}_{1}-\mathbf{x}_{2}$ etc. We define the quantity%
\begin{equation}
B_{AHS}(x)\equiv y_{AHS}(x)\left[  \varphi_{s}(x)-\varphi_{AHS}(x)\right]  ,
\label{blip}%
\end{equation}
in terms of the so-called cavity function $y_{AHS}(x)\equiv g_{AHS}%
(x)/\varphi_{AHS}(x)=\frac{2}{\rho^{2}}\frac{\delta\mathcal{A}}{\delta
\varphi(x)}$ and radial distribution function $g_{AHS}(x)$ pertaining to an
appropriate AHS potential which is the reference state. Both these functions
depend on $\rho$, $T$ and the effective adhesive parameter $\tau$, the latter
to be determined variationally. From now on, we omit the subscript $AHS$ in
$B_{AHS}(x)$, $g_{AHS}(x)$ etc. for the sake of brevity.

We next choose $\tau$ by requiring that the first-order correction to the
excess free energy vanishes%
\begin{equation}
\int d\mathbf{x}\;B(x)=0. \label{EIS}%
\end{equation}
This is the analogue of Eq. (\ref{vrwma}). Hence, in the spirit of the
previous section, we split up this integral into two parts. The first
indicates that the tail of the electrostatic interaction is compensated by
part of the original square well attraction%
\begin{equation}
\int_{2+\epsilon}^{\infty}dx\;x^{2}B(x)=0 \label{dem1}%
\end{equation}
($0<\varepsilon\leq\delta$) and yields $\epsilon$. The second determines the
density dependent strength $\tau$ of the AHS interaction%
\begin{equation}
\int_{2}^{2+\epsilon}dx\;x^{2}B(x)=0. \label{dem2}%
\end{equation}
This expresses the fact that the reference potential has to compensate for the
remaining part of the original interaction.

\subsection*{\textit{2. Approximate radial distribution function for the
Baxter potential}}

In order to be able to determine $\tau$ from Eqs. (\ref{dem1}) and
(\ref{dem2}), we need to know $g(x)$, the radial distribution function of the
reference interaction, the AHS potential. In the Percus-Yevick approximation
developed by Baxter, $g(x)$ has a singular contribution $g_{\omega}(x)$ which,
after the limit $\omega\rightarrow0$, acts like a delta function and results
from the stickiness of the interaction at the surfaces of two touching
spheres. We split $g(x)$ into $g_{\omega}(x)$ and a regular term
$\widetilde{g}(x)$ \cite{BAX}%

\begin{equation}
g(x)=\widetilde{g}(x)+g_{\omega}(x) \label{rad}%
\end{equation}
with%
\begin{equation}
g_{\omega}(x)=\left\{
\begin{array}
[c]{ll}%
0 & x<2\\
\frac{\lambda(2+\omega)}{12\omega}+O(1) & 2\leq x\leq2+\omega\\
0 & x>2+\omega
\end{array}
\right.
\end{equation}
analogously to Eq. (\ref{AHS}), where the amplitude $\lambda$ is the smaller
of the two solutions of%
\begin{equation}
\tau=\frac{1+\eta/2}{(1-\eta)^{2}}\frac{1}{\lambda}-\frac{\eta}{1-\eta}%
+\frac{\eta}{12}\lambda. \label{lambda}%
\end{equation}

For $x\,<2$, $\widetilde{g}(x)$ equals zero owing to the hard-core repulsion,
whereas $\widetilde{g}(x)$ tends to unity for large $x$. For proteins, it
turns out that $\varphi_{s}(x)-\varphi_{AHS}(x)$ is often appreciably nonzero
only near the surface of the sphere so we approximate $\widetilde{g}(x)$ in
the interval $2\leq x\leq4$ by the first two terms of its Taylor expansion%
\begin{equation}
\widetilde{g}(x)\simeq\left\{
\begin{array}
[c]{ll}%
0 & x<2\\
G(1+H(x-2)) & 2\leq x\leq4\\
1 & x>4
\end{array}
\right.  , \label{rad1}%
\end{equation}
The two constants have been derived by Bravo Yuste and Santos \cite{BRA}%
\begin{equation}
G=\lambda\tau\label{A}%
\end{equation}
and%
\begin{equation}
H=\frac{\eta}{2\tau(1-\eta)}\left(  \frac{\eta(1-\eta)}{12}\lambda^{2}%
-\frac{1+11\eta}{12}\lambda+\frac{1+5\eta}{1-\eta}-\frac{9(1+\eta)}%
{2(1-\eta)^{2}}\frac{1}{\lambda}\right)  .
\end{equation}
Numerical work \cite{KRA} bears out that Eq. (\ref{rad1}) is quite reasonable
since the range of both attractive and electrostatic forces is much smaller
than the diameter of the protein.

\subsection*{\textit{3. Determination of the effective adhesion}}

We next determine $\tau$ from Eq. (\ref{dem2}), first using Eq. (\ref{dem1})
to obtain $\epsilon$. From Eqs. (\ref{AHS}), (\ref{blip}) and (\ref{rad}), the
function $B(x)$ can be shown to have the following form (repressing terms that
ultimately disappear in the limit $\omega\rightarrow0$)%
\begin{equation}
B(x)=\widetilde{B}(x)-g_{\omega}(x), \label{blip2}%
\end{equation}
where the regular term is given by%
\begin{equation}
\widetilde{B}(x)=\left\{
\begin{array}
[c]{ll}%
0 & 0\leq x\leq2\\
(e^{-U(x)}-1)\widetilde{g}(x) & x>2
\end{array}
\right.  .
\end{equation}
Eq. (\ref{dem1}) may be conveniently expressed as%
\begin{equation}
\int_{2+\epsilon}^{\infty}dx\;x^{2}B(x)=\int_{2+\epsilon}^{2+\delta}%
dx\;x^{2}\widetilde{B}(x)+\int_{2+\delta}^{\infty}dx\;x^{2}\widetilde{B}(x)=0.
\end{equation}
Using $\int_{2+\epsilon}^{2+\delta}dx\;f(x)\simeq\frac{1}{2}(\delta
-\epsilon)\left[  f(2+\delta)+f(2+\epsilon)\right]  $ and neglecting terms of
order $\delta^{2}$ and $\epsilon^{2}$, we write the first integral as%
\begin{equation}
\int_{2+\epsilon}^{2+\delta}dx\;x^{2}B(x)\simeq G(\delta-\epsilon)K_{1}%
(\delta,\epsilon), \label{eps1}%
\end{equation}
with%
\begin{equation}
K_{1}(\delta,\epsilon)\equiv2\left(  e^{U_{A}}e^{-\frac{\xi}{1+\delta
/2}e^{-\mu\delta}}-1\right)  \left(  1+\left(  1+H\right)  \delta\right)
+2\left(  e^{U_{A}}e^{-\frac{\xi}{1+\epsilon/2}e^{-\mu\epsilon}}-1\right)
\left(  1+\left(  1+H\right)  \epsilon\right)  .
\end{equation}
Again, we stress that, although $\delta\ll1$ and $\epsilon\ll1$, $\mu\delta$
and $\mu\epsilon$ may be of order unity. Furthermore, we note that if we take
the limit $\eta\downarrow0$, then $\lambda\rightarrow\tau^{-1}$ and
$G\rightarrow1$, so we recover Eq. (\ref{eps}) if we neglect terms of order
$\delta$ and $\epsilon$. We tackle the second integral by adopting the
approximation: $1-\exp(-U(x))=1-\exp(2\xi x^{-1}e^{-\mu(x-2)})\simeq2\xi
x^{-1}e^{-\mu(x-2)}-2\xi^{2}x^{-2}e^{-2\mu(x-2)}+2\xi^{3}x^{-2}e^{-3\mu
(x-2)}/3$ (note that in this Taylor expansion of the exponential we have
replaced one factor $x^{-1}$ by $2^{-1}$ in the last term). We then write%
\begin{equation}
-\int_{2+\delta}^{\infty}dx\;x^{2}B(x)\simeq G\left(  (1+\delta H)P_{1}%
+HP_{2}\right)
\end{equation}
with%
\begin{equation}
P_{1}=\int_{2+\delta}^{\infty}dx\;x^{2}(1-e^{-U(x)})\simeq\frac{8}{\mu^{2}%
}(1+\mu\delta)M+\frac{16}{\mu}M\left(  1-M+\frac{8}{9}M^{2}\right)
\end{equation}
and%
\begin{equation}
P_{2}=\int_{2+\delta}^{\infty}dx\;x^{2}(x-2-\delta)(1-e^{-U(x)})\simeq\frac
{8}{\mu^{3}}(2+\mu\delta)M+\frac{16}{\mu^{2}}\left(  M-\frac{1}{2}M^{2}%
+\frac{8}{27}M^{3}\right)  .
\end{equation}
Here, $M\equiv\xi e^{-\mu\delta}/4$. Using the approximations $1-M+8M/9\simeq
(1+M)^{-1}$ and $M-M^{2}/2+8M^{3}/27\simeq\log(1+M)$, we arrive at%
\begin{equation}
P_{1}\simeq\frac{8}{\mu^{2}}(1+\mu\delta)M+\frac{16}{\mu}\frac{M}{1+M}%
\end{equation}
and%
\begin{equation}
P_{2}\simeq\frac{8}{\mu^{3}}(2+\mu\delta)M+\frac{16}{\mu^{2}}\log(1+M).
\end{equation}
Hence, the variable $\epsilon$, which depends on the density by virtue of the
density dependence of $G$ and $H$, is determined iteratively from%
\begin{equation}
\delta-\epsilon_{new}=\frac{(1+\delta H)P_{1}+HP_{2}}{K_{1}(\delta
,\epsilon_{old})}. \label{vale}%
\end{equation}
One starts with $\epsilon_{old}=\delta$ and iterates until a stationary
$\epsilon_{new}$ is reached.

The next step is to calculate $\tau$ from Eq. (\ref{dem2}), which, with the
help of Eq. (\ref{blip2}), is equivalent to the expression%
\begin{equation}
\int_{2}^{2+\epsilon}dx\;x^{2}\widetilde{B}(x)=\frac{2\lambda}{3}.
\label{dett}%
\end{equation}
We have taken the limit $\omega\rightarrow0$. Again using the approximation
$\int_{2}^{2+\epsilon}dx~f(x)\simeq\frac{1}{2}\epsilon\left[  f(2+\epsilon
)+f(2)\right]  $, we write%
\begin{equation}
\int_{2}^{2+\epsilon}dx\;x^{2}\widetilde{B}(x)\simeq2G\epsilon\left[  \left(
e^{U_{A}}e^{-\frac{\xi}{1+\epsilon/2}e^{-\mu\epsilon}}-1\right)  \left(
1+\left(  1+H\right)  \epsilon\right)  +\left(  e^{U_{A}}e^{-\xi}-1\right)
\right]  .
\end{equation}
Together with the expressions Eq. (\ref{dett}) and $G=\lambda\tau$ (Eq.
(\ref{A})), this leads to%
\begin{equation}
\frac{1}{\tau}\simeq3\epsilon\left[  \left(  e^{U_{A}}e^{-\frac{\xi
}{1+\epsilon/2}e^{-\mu\epsilon}}-1\right)  \left(  1+\left(  1+H\right)
\epsilon\right)  +\left(  e^{U_{A}}e^{-\xi}-1\right)  \right]  . \label{valt}%
\end{equation}
Accordingly, $\tau$ may be determined iteratively if we recall that both $H$
and $\epsilon$ also depend on $\tau$. A way of quickly determining $\tau$ and
$\epsilon$ is choosing a starting value for both ($\epsilon=\delta$ and
$\tau=0.2$ say), and then alternately using Eqs. (\ref{vale}) and (\ref{valt})
until the iterates become stationary.

\subsection*{B. Application to lysozyme}

We have already determined the interaction in section IIB2b. We next compute
$\tau$ iteratively and it now depends on both the density of protein and the
ionic strength. (See table III).

Thermodynamic properties like the osmotic compressibility $\kappa_{T}$ are
also simply obtained from $\tau$. In the Percus-Yevick approximation,
$\kappa_{T}$ is given by \cite{BAX}%

\begin{equation}
\left(  \rho k_{B}T\kappa_{T}\right)  ^{-1}\equiv\frac{1}{k_{B}T}%
\frac{\partial\Pi}{\partial\rho}=\frac{(1+2\eta-\lambda\eta(1-\eta))^{2}%
}{(1-\eta)^{4}}, \label{compr}%
\end{equation}
where $\lambda$ is the smaller of the two solutions of Eq. (\ref{lambda}).
Fig. 8 compares the predicted density dependence of the (scaled) inverse
osmotic compressibility at various ionic strengths with experimental data from
\cite{PIA2} and \cite{ROS2}.

\section*{IV. DISCUSSION}

One difficulty in comparing our computations with experiment has been the
substantial margin of error in the osmotic measurements. In the case of other
biomacromolecules like rodlike DNA, it has been possible to obtain the second
virial $B_{2}$ at better than $10\%$ accuracy \cite{NIC,FER,WIS}. One
possibility for the occurrence of discrepancies in $B_{2}$ is the variety of
lysozyme types. Poznanski et al \cite{POZ} have established that popular
commercial lysozyme preparations like Seikagaku and Sigma exhibit significant
differences under dynamic light scattering. Nevertheless, the variation in
$B_{2}$ at, say, about $0.5$ M NaCl (see Fig. 3), is so large that it needs to
be explained.

The relatively large variation in the experimental measurements of $B_{2}$
makes it difficult to falsify stringently other models of attractive forces
like that of van der Waals type, for instance. It proves feasible to get
satisfactory agreement with the experimental data displayed in Fig. 3 if we
let the dispersion interaction be given by the nonretarded Hamaker potential
\cite{VER} for spheres of dimensions appropriate for lysozyme, with an
adjustable Hamaker constant of order $k_{B}T$ though with a very short cut-off
at around $0.1-0.2$ $%
\operatorname{nm}%
$. However, the necessity of such a cut-off, which is already beyond the limit
of validity of continuum approximations, may be viewed as positing the
equivalent of a short-range interaction like that of Eq. (\ref{pot}), in large
part. The long-range dispersion interaction beyond some distance much smaller
than the radius $a$, plays only a minor role.

Stell \cite{STE} has criticized the Baxter limit because divergences in the
free energy appear at the level of the $12$th virial. Therefore, the most
straightforward way to interpret our liquid state theory is to stress that our
zero-order theory describes the reference state only up to and including the
$11$th virial within the Percus-Yevick approximation. The analysis of phase
transitions must be viewed with caution (for a comparison of recent
simulations---taking the limit of zero polydispersity after the limit of
vanishing well depth---with Percus-Yevick theory, see \cite{MIL}). A second
problem is here that, at large ionic strengths, a considerable electrostatic
repulsion is balanced against a significant attraction (see Fig. 1) and it is
difficult to see how good such a compensatory scheme should work at high
concentrations near dense packing.

In summary, we have presented a fairly good theory of the ionic-strength
dependence of the osmotic properties of lysozyme in terms of a sticky
interaction which is independent of charge or salt concentration. This
conclusion, by itself, is not new for it has been reached earlier by
formulating numerical work incorporating short-range forces and screened
electrostatics and comparing it with X-ray scattering \cite{MAL,TAR} and
liquid-liquid phase separation \cite{VLA,GRI,PET}. The merit of the current
analysis is its transparency because it is analytical and it is based on a
nonperturbative variational principle for general short-range potentials so it
may be readily generalized.

\section*{V. APPENDIX}

\subsection*{A. Effective charge}

For the repulsive tail of the two particle interaction, we use the
Debye-H\"{u}ckel potential, which is the far-field solution of the
Poisson-Boltzmann equation. In our case, the (dimensionless) potential at the
surface is often merely of order unity, so the Debye-H\"{u}ckel potential
slightly overestimates the solution to the Poisson-Boltzmann equation. To
remedy this, we use a renormalized charge within the Debye-H\"{u}ckel
potential, chosen in such a way that, at large distances, the Debye-H\"{u}ckel
potential coincides with the tail of the solution of the Poisson-Boltzmann
equation determined by the real charge \cite{GRO}. This will result in an
underestimation of the potential at small separations, but the form of the
Debye-H\"{u}ckel potential we use here (Eq. (\ref{debye})) is in fact only
accurate at large separations and overestimates the interaction at small
separations appreciably i.e. when overlap of the two double layers occurs (by
about 20\%, see \cite{VER}). The two effects thus partly cancel, although the
latter effect is larger than the former.

The Poisson-Boltzmann equation for the dimensionless potential $\psi\left(
r\right)  =q\phi(r)/k_{B}T$ of a single sphere of radius $a$ and total charge
$qZ$, assumed positive for convenience, immersed in a solvent with Bjerrum
length $Q$, at a concentration of ions leading to a Debye length $\kappa$, is
written as%

\begin{equation}
\frac{1}{r^{2}}\frac{d}{dr}r^{2}\frac{d}{dr}\psi\left(  r\right)  =\kappa
^{2}\sinh\psi\left(  r\right)  , \label{PB}%
\end{equation}
with boundary conditions%
\begin{equation}
\left.  \frac{d}{dr}\psi\left(  r\right)  \right\vert _{r=a}=\frac{ZQ}{a^{2}%
};\;\lim_{r\rightarrow\infty}\psi(r)=0.
\end{equation}
Linearizing Eq. (\ref{PB}) ($\psi\ll1$), we find the Debye-H\"{u}ckel solution%
\begin{equation}
\psi_{0}=\frac{ZQ}{1+\mu}\frac{e^{-\kappa(r-a)}}{r}.
\end{equation}
We next derive the first-order correction to this solution. Putting
$\psi\left(  r\right)  =\psi_{0}\left(  r\right)  +\psi_{1}\left(  r\right)
$, with $\left\vert \psi_{1}\left(  r\right)  \right\vert \ll\left\vert
\psi_{0}\left(  r\right)  \right\vert $, results in the following linear
differential equation for $\psi_{1}$%
\begin{equation}
\frac{1}{r^{2}}\frac{d}{dr}r^{2}\frac{d}{dr}\psi_{1}\left(  r\right)
=\frac{1}{6}\kappa^{2}\psi_{0}^{3}\left(  r\right)  .
\end{equation}
Keeping in mind that $\psi_{1}\left(  r\right)  =o(\psi_{0}\left(  r\right)
)$, as $r\rightarrow\infty$, we integrate the differential equation once to
obtain%
\begin{equation}
\frac{d}{dr}\psi_{1}\left(  r\right)  =-\frac{\kappa^{2}}{6}\left(
\frac{ZQe^{\mu}}{1+\mu}\right)  ^{3}\frac{E_{1}(3\kappa r)}{r^{2}}%
\end{equation}
and a second time to derive%
\begin{equation}
\psi_{1}\left(  r\right)  =-\frac{\kappa^{3}}{6}\left(  \frac{ZQe^{\mu}}%
{1+\mu}\right)  ^{3}\left(  \frac{e^{-3\kappa r}}{\kappa r}-\left(  3+\frac
{1}{\kappa r}\right)  E_{1}(3\kappa r)\right)  ,
\end{equation}
where $E_{1}(x)$ is the exponential integral defined by $E_{1}(x)=\int
_{x}^{\infty}dt\,t^{-1}e^{-t}$. Using the first of the two boundary
conditions, we then determine the renormalized charge $Z_{eff}$%
\begin{align}
Z_{eff}  &  =\frac{a^{2}}{Q}\left.  \frac{d}{dr}\psi\left(  r\right)
\right\vert _{r=a}=\frac{a^{2}}{Q}\left.  \frac{d}{dr}\psi_{0}\left(
r\right)  \right\vert _{r=a}+\frac{a^{2}}{Q}\left.  \frac{d}{dr}\psi
_{1}\left(  r\right)  \right\vert _{r=a}=\nonumber\\
&  =Z-\frac{\mu}{18}\left(  \frac{Q}{a}\right)  ^{2}\left(  \frac{Z}{1+\mu
}\right)  ^{3}F(\mu), \label{cha}%
\end{align}
where%
\begin{equation}
F(\mu)\equiv3\mu e^{3\mu}E_{1}(3\mu)\sim1-\frac{1}{3\mu}+\frac{2}{9\mu^{2}%
}-\ldots
\end{equation}

Recapitulating, we have calculated, to leading order, the charge $Z_{eff}$
which has to be inserted into the Debye-H\"{u}ckel potential (Eq.
(\ref{debye})) so that this has the correct asymptotic behavior at large $r$,
coinciding with the tail of the Poisson-Boltzmann solution.

\subsection*{B. Dependence of $B_{2}$ on ionic strength}

Here, we prove some simple inequalities describing the behavior of the second
virial coefficient as a function of the ionic strength for an interaction
consisting of a Debye-H\"{u}ckel repulsion $U_{DH}(x)$ and a general
attractive potential $U_{A}(x)$, the latter not depending on the ionic
strength. If we let $U(x)=U_{DH}(x)+U_{A}(x)$, then $B_{2}$ is given by Eq.
(\ref{vir2}) with%
\begin{equation}
J=\int_{2}^{\infty}dx\;x^{2}\left(  1-e^{-U\left(  x\right)  }\right)  .
\end{equation}
Then, we have%
\begin{equation}
\frac{dJ}{d\mu}=\int_{2}^{\infty}dx\;x^{2}\frac{dU_{DH}\left(  x\right)
}{d\mu}e^{-U\left(  x\right)  }=\int_{2}^{\infty}dx\;x^{2}\left(  \frac
{d\ln\xi}{d\mu}-(x-2)\right)  U_{DH}\left(  x\right)  \,e^{-U\left(  x\right)
}.
\end{equation}
In Fig. 9 we see that in the regime of interest $\frac{d\ln\xi}{d\mu}<0$, so
we conclude that%
\begin{equation}
\frac{dB_{2}}{d\mu}=\frac{3}{8}B_{2}^{HS}\frac{dJ}{d\mu}<0.
\end{equation}
In the same way it is clear from the second derivative%
\begin{equation}
\frac{d^{2}J}{d\mu^{2}}=\int_{2}^{\infty}dx\;x^{2}\left(  \frac{d^{2}\ln\xi
}{d\mu^{2}}+\left(  \frac{d\ln\xi}{d\mu}-(x-2)\right)  ^{2}\left(
1-U_{DH}\left(  x\right)  \right)  \right)  \,U_{DH}\left(  x\right)
\,e^{-U\left(  x\right)  }%
\end{equation}
and the fact that $\frac{d^{2}\ln\xi}{d\mu^{2}}\gtrsim0$ in the regime of
interest, that%
\begin{equation}
\frac{d^{2}B_{2}}{d\mu^{2}}=\frac{3}{8}B_{2}^{HS}\frac{d^{2}J}{d\mu^{2}}>0,
\end{equation}
if $U_{DH}\left(  2\right)  <1$, i.e. if $\xi<1$ (a sufficient condition).

\subsection*{C. Corrections to the free energy}

In section III, we viewed a suspension of proteins as a system of spheres with
an AHS interaction and we chose the parameter $\tau$ of the AHS potential such
that the first order correction in the functional expansion of the free energy
(Eq. (\ref{WCA})) vanishes (see Eq. (\ref{EIS})). In an attempt to justify
this approximation and explore its regime of applicability, we estimate the
size of the second order correction to the free energy (from Eq. (\ref{WCA}))
which is either positive or negative definite%

\begin{equation}
\Delta\equiv\frac{\eta^{2}}{2}\left(  \frac{3}{4\pi}\right)  ^{2}a^{3}%
V^{-1}\int d\mathbf{x}_{1}d\mathbf{x}_{2}d\mathbf{x}_{3}\;B(x_{12}%
)B(x_{13})h(x_{23})=\frac{9}{4}\eta^{2}Y. \label{WCAc}%
\end{equation}
It is convenient to rewrite the integral in such a way that the angular
integration can be performed explicitly (see below).%
\begin{align}
Y  &  \equiv\int_{0}^{\infty}dt\;t^{2}B(t)\int_{0}^{\infty}ds\;s^{2}%
B(s)\int_{0}^{\pi}d\vartheta\;\sin\vartheta\;h(\sqrt{s^{2}+t^{2}%
-2st\cos\vartheta})=\\
&  =2\int_{0}^{\infty}dt\;tB(t)\int_{t}^{\infty}ds\;sB(s)\int_{s-t}%
^{s+t}du\;uh(u).\nonumber
\end{align}
Here we have used the Kirkwood superposition approximation $J_{BM}%
^{(3)}(\mathbf{x}_{1},\mathbf{x}_{2},\mathbf{x}_{3})=h(x_{23})$ \cite{AND},
where $h(x)=g(x)-1$ is the pair correlation function. We have employed the
substitution $u^{2}=s^{2}+t^{2}-2st\cos\vartheta$, with $\vartheta$ the angle
between $\mathbf{x}_{12}$ and $\mathbf{x}_{13}$. Using the expression for
$g(x)$ (Eq. (\ref{rad})) and defining $\widetilde{h}(x)=\widetilde{g}(x)-1$,
we split $Y$ into three parts
\begin{equation}
Y=Y_{0}+Y_{1}+Y_{2}%
\end{equation}
where we have introduced the limit $\omega\rightarrow0$ and where%
\begin{align}
Y_{0}  &  \equiv\frac{2\lambda}{3}\int_{2}^{\infty}dt\;tB(t)\int_{t}%
^{t+2}ds\;sB(s)\simeq\\
&  \simeq\frac{2\lambda}{3}\int_{2}^{\infty}dt\;tB(t)\int_{t}^{\infty
}ds\;sB(s)=\frac{\lambda}{3}\left[  \int_{2}^{\infty}dt\;tB(t)\right]
^{2},\nonumber
\end{align}%
\begin{equation}
Y_{1}\equiv2\int_{2}^{\infty}dt\;tB(t)\int_{t}^{t+2}ds\;sB(s)\int_{s-t}%
^{s+t}du\;u\widetilde{h}(u). \label{SPL}%
\end{equation}
and%
\begin{equation}
Y_{2}\equiv2\int_{2}^{\infty}dt\;tB(t)\int_{t+2}^{t+4}ds\;sB(s)\int
_{s-t}^{s+t}du\;u\widetilde{h}(u)\ll Y_{1}%
\end{equation}
To simplify Eq. (\ref{SPL}), we substitute Eq. (\ref{rad1}) and note that
$s+t\geq4$ and $0\leq s-t\leq2$. We then derive%
\begin{equation}
\int_{s-t}^{s+t}du\;u\widetilde{h}(u)=\frac{2}{3}\left(  9G+10GH-12\right)
+\frac{1}{2}(s-t)^{2}.
\end{equation}
Next, using Eq. (\ref{EIS}), we integrate the nonconstant term leading to a
product of two integrals%
\begin{align}
\int_{2}^{\infty}dt\;tB(t)\int_{t}^{t+2}ds\;sB(s)(s-t)^{2}  &  \simeq\int
_{2}^{\infty}dt\;tB(t)\int_{t}^{\infty}ds\;sB(s)(s-t)^{2}=\\
&  =\left[  \int_{2}^{\infty}dt\;tB(t)\right]  \left[  \int_{2}^{\infty
}ds\;s^{3}B(s)\right]  .\nonumber
\end{align}
Hence, $Y_{1}$ is written in terms of one-dimensional integrals%
\begin{equation}
Y_{1}\simeq\frac{2}{3}\left(  9G+10GH-12\right)  \left[  \int_{2}^{\infty
}dt\;tB(t)\right]  ^{2}+\left[  \int_{2}^{\infty}dt\;tB(t)\right]  \left[
\int_{2}^{\infty}ds\;s^{3}B(s)\right]  ,
\end{equation}
and this is also the case for $Y$%
\begin{equation}
Y\simeq\frac{2}{3}\left(  9G+10GH-12+\frac{\lambda}{2}\right)  \left[
\int_{2}^{\infty}dt\;tB(t)\right]  ^{2}+\left[  \int_{2}^{\infty
}dt\;tB(t)\right]  \left[  \int_{2}^{\infty}ds\;s^{3}B(s)\right]  .
\label{int1}%
\end{equation}

Our goal is to obtain explicit approximations for these integrals by
expediently using Eqs. (\ref{dem1}) and (\ref{dem2}). First, we consider
integrals on the interval $[2,2+\epsilon]$ which are dominated by the singular
part of $B(x)$. We substitute Eq. (\ref{blip2}) into (\ref{dem2}) and let
$\omega\rightarrow0$%
\begin{equation}
\int_{2}^{2+\epsilon}dt\;t^{2}\widetilde{B}(t)=\frac{2\lambda}{3}. \label{kpl}%
\end{equation}
We use this relation to rewrite part of one of the integrals in Eq.
(\ref{int1}) in two ways, noting that $\epsilon\ll1$.%
\begin{equation}
\int_{2}^{2+\epsilon}dt\;tB(t)=-\frac{\lambda}{3}+\int_{2}^{2+\epsilon
}dt\;t\widetilde{B}(t)=-\frac{1}{2}\int_{2}^{2+\epsilon}dt\;(t-2)t\widetilde
{B}(t)\simeq-\frac{\epsilon}{4}\int_{2}^{2+\epsilon}dt\;t\widetilde{B}(t).
\label{tgw}%
\end{equation}
We thus conclude that%
\begin{equation}
\int_{2}^{2+\epsilon}dt\;t\widetilde{B}(t)\simeq\left(  1-\frac{\epsilon}%
{4}\right)  \frac{\lambda}{3}%
\end{equation}
so the first equality in Eq. (\ref{tgw}) allows us to attain the explicit
expression%
\begin{equation}
\int_{2}^{2+\epsilon}dt\;tB(t)\simeq-\frac{\lambda\epsilon}{12}. \label{t1}%
\end{equation}
Similarly, we use Eqs. (\ref{blip2}) and (\ref{kpl}) to evaluate part of the
other integral in Eq. (\ref{int1}).%
\begin{equation}
\int_{2}^{2+\epsilon}dt\;t^{3}B(t)=-\frac{4\lambda}{3}+\int_{2}^{2+\epsilon
}dt\;t^{3}\widetilde{B}(t)=\int_{2}^{2+\epsilon}dt\;(t-2)t^{2}\widetilde
{B}(t)\simeq\frac{\lambda\epsilon}{3}. \label{t2}%
\end{equation}
We note that both integrals in Eqs. (\ref{t1}) and (\ref{t2}) are
$O(\epsilon)$ because the integral in Eq. (\ref{kpl}) is independent of
$\epsilon$ owing to the singular part of $B(x)$. If $B(x)$ had been completely
regular, the integrals in Eqs. (\ref{t1}) and (\ref{t2}) would have been
$O(\epsilon^{2})$.

We next consider the remaining two integrals on the interval $[2+\epsilon
,\infty)$. We start by splitting Eq. (\ref{dem1}) into two parts since
$2+\delta$ demarcates two different regimes%
\begin{equation}
\int_{2+\epsilon}^{2+\delta}dt\;t^{2}B(t)+\int_{2+\delta}^{\infty}%
dt\;t^{2}B(t)=0.
\end{equation}
Using this equation and the approximation $B(t)\simeq-2\xi e^{-\mu(t-2)}/t$,
we may simplify the two integrals, ultimately omitting $O(\delta)$ terms%
\begin{align}
\int_{2+\epsilon}^{\infty}dt\;tB(t)  &  =\int_{2+\epsilon}^{2+\delta
}dt\;tB(t)+\int_{2+\delta}^{\infty}dt\;tB(t)\simeq\nonumber\\
&  \simeq\frac{1}{2}\left(  1-\frac{\delta}{2}\right)  \int_{2+\epsilon
}^{2+\delta}dt\;t^{2}B(t)+\int_{2+\delta}^{\infty}dt\;tB(t)=\nonumber\\
&  =\frac{\delta}{4}\int_{2+\delta}^{\infty}dt\;t^{2}B(t)-\frac{1}{2}%
\int_{2+\delta}^{\infty}dt\;t(t-2)B(t)\simeq\frac{\xi}{\mu^{2}}e^{-\mu\delta}
\label{t3}%
\end{align}%
\begin{align}
\int_{2+\epsilon}^{\infty}dt\;t^{3}B(t)  &  =\int_{2+\epsilon}^{2+\delta
}dt\;t^{3}B(t)+\int_{2+\delta}^{\infty}dt\;t^{3}B(t)=\nonumber\\
&  \simeq(2+\delta)\int_{2+\epsilon}^{2+\delta}dt\;t^{2}B(t)+\int_{2+\delta
}^{\infty}dt\;t^{3}B(t)=\nonumber\\
&  =-\delta\int_{2+\delta}^{\infty}dt\;t^{2}B(t)+\int_{2+\delta}^{\infty
}dt\;t^{2}(t-2)B(t)\simeq-4\frac{\xi}{\mu^{2}}e^{-\mu\delta}. \label{t4}%
\end{align}
We remark that both expressions in Eqs. (\ref{t3}) and (\ref{t4}) are
$O(\mu^{-2})$ because $B(t)$ is regular for $t\geq2+\epsilon$. We then combine
Eqs. (\ref{t1}) and (\ref{t3}), and Eqs. (\ref{t2}) and (\ref{t4})%
\begin{equation}
\int_{2}^{\infty}dt\;tB(t)\simeq-\frac{\lambda\epsilon}{12}+\frac{\xi}{\mu
^{2}}e^{-\mu\delta}\simeq-\frac{1}{4}\int_{2}^{\infty}ds\;s^{3}B(s).
\label{prt3}%
\end{equation}
Finally, using Eqs. (\ref{WCAc}), (\ref{int1}) and (\ref{prt3}), we arrive at
an approximation for the correction to the free energy%
\begin{equation}
\Delta=\frac{9}{4}\eta^{2}Y\simeq\frac{9}{2}\eta^{2}\left(  G+H-6+\frac
{\lambda}{6}\right)  \left[  \frac{\xi}{\mu^{2}}e^{-\mu\delta}-\frac
{\lambda\epsilon}{12}\right]  ^{2}. \label{cor}%
\end{equation}
Despite the variety of approximations used, this expression still retains its
\textquotedblleft definite\textquotedblright\ character (it turns out to be
negative in the numerical calculations below). However, the numerical
coefficients within the last quadratic factor are not exact. Furthermore, the
status of the present theory differs from that of the Weeks-Chandler-Anderson
theory \cite{AND}. In the latter, $\Delta$ is of fourth order in the
perturbation whereas it is basically quadratic here for the reason stated
below Eq. (\ref{t2}).

To estimate the importance of this correction, we first calculate the osmotic
pressure resulting from the neglect of second and higher order terms in the
functional expansion Eq. (\ref{WCA}). This amounts to determining $\tau$ from
Eqs. (\ref{lambda}), (\ref{vale}) and (\ref{valt}) and then computing the
osmotic pressure from Ref. \cite{BAX}%
\begin{equation}
\frac{\Pi}{\rho k_{B}T}=\frac{1+\eta+\eta^{2}-\lambda\eta(1-\eta)(1+\frac
{1}{2}\eta)+\lambda^{3}\eta^{2}(1-\eta)^{3}/36}{(1-\eta)^{3}}.\label{osmpr}%
\end{equation}
Then, we evaluate the correction to the osmotic pressure due to the second
order term in Eq. (\ref{WCA}). The osmotic pressure is related to the free
energy by%
\begin{equation}
\frac{\Pi}{\rho k_{B}T}=-\eta\frac{\partial(\rho^{-1}\mathcal{A})}%
{\partial\eta}.\label{osm}%
\end{equation}
Because $Y$ depends only weakly on $\eta$, we approximate the correction to
the osmotic pressure by%
\begin{equation}
-\eta\frac{\partial\Delta}{\partial\eta}\simeq-2\Delta.\label{crr}%
\end{equation}
We have compiled the pressure and its correction in table IV for the same sets
of parameters as in table I (omitting the trivial case where $\eta=0$).

\section*{Tables}

\hspace{0.3in}%
\begin{tabular}
[c]{|l|l|l|l|l|l|l|l|l|l|l|}\hline
$I$ (M) & $0.05$ & $0.1$ & $0.15$ & $0.2$ & $0.25$ & $0.3$ & $0.45$ & $1$ &
$1.5$ & $2$\\\hline
$Z$ & $9.5$ & $9.8$ & $10.0$ & $10.1$ & $10.2$ & $10.2$ & $10.3$ & $10.4$ &
$10.4$ & $10.4$\\\hline
$Z_{eff}$ & $8.8$ & $9.2$ & $9.4$ & $9.6$ & $9.7$ & $9.8$ & $10.0$ & $10.2$ &
$10.3$ & $10.3$\\\hline
$\overline{Z}$ & $7.8$ & $8.2$ & $8.4$ & $8.6$ & $8.7$ & $8.8$ & $9.0$ & $9.2$
& $9.3$ & $9.3$\\\hline
$\xi$ & $2.52$ & $1.84$ & $1.48$ & $1.27$ & $1.10$ & $0.984$ & $0.752$ &
$0.409$ & $0.295$ & $0.229$\\\hline
$\mu$ & $1.25$ & $1.76$ & $2.16$ & $2.50$ & $2.79$ & $3.06$ & $3.74$ & $5.58$
& $6.83$ & $7.89$\\\hline
$\epsilon_{0}$ &  &  & $0.0208$ & $0.0466$ & $0.0585$ & $0.0644$ & $0.0720$ &
$0.0773$ & $0.0782$ & $0.0785$\\\hline
$U_{S}$ &  &  & $2.26$ & $2.52$ & $2.70$ & $2.82$ & $3.05$ & $3.37$ & $3.47$ &
$3.53$\\\hline
$\tau$ &  &  & $0.933$ & $0.314$ & $0.205$ & $0.164$ & $0.115$ & $0.0767$ &
$0.0684$ & $0.0642$\\\hline
\end{tabular}
\smallskip

\textbf{Table I:} Values of the actual charge $Z$ (from \cite{KUE}), the
renormalized or effective charge $Z_{eff}$ (from Eq. (\ref{cha})), the lowered
effective charge $\overline{Z}=Z_{eff}-1$, and dimensionless interaction
parameters $\xi$ and $\mu$, and $\epsilon_{0}$, $U_{S}$ and $\tau$ as a
function of the ionic strength $I$. The pH equals $4.5$ and $\xi$ has been
calculated using the lowered effective charge $\overline{Z}$. Values of
$U_{S}$ and $\tau$ have been computed using Eqs. (\ref{detus}) and
(\ref{dettau}), respectively, and $\epsilon_{0}$ has been calculated using the
procedure described immediately after Eq. (\ref{deteps}).\newpage

\hspace{0.3in}%
\begin{tabular}
[c]{|l|l|l|l|l|l|l|l|l|l|l|}\hline
$I$ (M) & $0.05$ & $0.1$ & $0.15$ & $0.2$ & $0.25$ & $0.3$ & $0.45$ & $1$ &
$1.5$ & $2$\\\hline
$Z$ & $6.9$ & $7.0$ & $7.1$ & $7.2$ & $7.2$ & $7.3$ & $7.3$ & $7.1$ & $6.9$ &
$6.8$\\\hline
$Z_{eff}$ & $6.6$ & $6.8$ & $6.9$ & $7.0$ & $7.0$ & $7.1$ & $7.2$ & $7.0$ &
$6.9$ & $6.8$\\\hline
$\overline{Z}$ & $5.6$ & $5.8$ & $5.9$ & $6.0$ & $6.0$ & $6.1$ & $6.2$ & $6.0$
& $5.9$ & $5.8$\\\hline
$\xi$ & $1.3$ & $0.920$ & $0.728$ & $0.616$ & $0.524$ & $0.473$ & $0.357$ &
$0.174$ & $0.119$ & $0.0889$\\\hline
$\mu$ & $1.25$ & $1.76$ & $2.16$ & $2.50$ & $2.79$ & $3.06$ & $3.74$ & $5.58$
& $6.83$ & $7.89$\\\hline
$\epsilon_{0}$ &  & $0.0493$ & $0.0640$ & $0.0695$ & $0.0725$ & $0.0741$ &
$0.0764$ & $0.0784$ & $0.0787$ & $0.0788$\\\hline
$U_{S}$ &  & $2.83$ & $3.03$ & $3.14$ & $3.23$ & $3.28$ & $3.39$ & $3.56$ &
$3.61$ & $3.63$\\\hline
$\tau$ &  & $0.212$ & $0.132$ & $0.108$ & $0.0943$ & $0.0877$ & $0.0758$ &
$0.0623$ & $0.0590$ & $0.0574$\\\hline
\end{tabular}
\smallskip

\textbf{Table II:} Same as table I, but now with a pH equal to $7.5$.\newpage

\hspace{0.3in}%
\begin{tabular}
[c]{|l|l|l|l|l|l|l|l|l|l|}\hline
$\eta$ &  & $0.15\text{ M}$ & $0.2\text{ M}$ & $0.25\text{ M}$ & $0.3\text{
M}$ & $0.45\text{ M}$ & $1\text{ M}$ & $1.5\text{ M}$ & $2\text{ M}$\\\hline
$0$ & $\tau$ & $0.829$ & $0.295$ & $0.194$ & $0.156$ & $0.110$ & $0.0735$ &
$0.0656$ & $0.0616$\\\hline
& $\varepsilon$ & $0.0230$ & $0.0483$ & $0.0596$ & $0.0653$ & $0.0725$ &
$0.0775$ & $0.0782$ & $0.0786$\\\hline
$0.05$ & $\tau$ & $0.712$ & $0.289$ & $0.193$ & $0.155$ & $0.110$ &  &  &
\\\hline
& $\varepsilon$ & $0.0266$ & $0.0492$ & $0.0600$ & $0.0655$ & $0.0725$ &  &  &
\\\hline
$0.1$ & $\tau$ & $0.620$ & $0.283$ & $0.192$ & $0.155$ & $0.110$ &  &  &
\\\hline
& $\varepsilon$ & $0.0303$ & $0.0502$ & $0.0603$ & $0.0656$ & $0.0725$ &  &  &
\\\hline
$0.15$ & $\tau$ & $0.544$ & $0.276$ & $0.191$ & $0.155$ & $0.110$ &  &  &
\\\hline
& $\varepsilon$ & $0.0342$ & $0.0514$ & $0.0607$ & $0.0657$ & $0.0724$ &  &  &
\\\hline
$0.2$ & $\tau$ & $0.482$ & $0.268$ & $0.190$ & $0.155$ & $0.110$ &  &  &
\\\hline
& $\varepsilon$ & $0.0383$ & $0.0528$ & $0.0611$ & $0.0658$ & $0.0723$ &  &  &
\\\hline
$0.3$ & $\tau$ & $0.380$ & $0.251$ & $0.186$ & $0.154$ & $0.110$ &  &  &
\\\hline
& $\varepsilon$ & $0.0477$ & $0.0563$ & $0.0624$ & $0.0663$ & $0.0722$ &  &  &
\\\hline
$0.4$ & $\tau$ & $0.300$ & $0.228$ & $0.179$ & $0.152$ & $0.110$ &  &  &
\\\hline
& $\varepsilon$ & $0.0600$ & $0.0619$ & $0.0651$ & $0.0677$ & $0.0724$ &  &  &
\\\hline
\end{tabular}
\smallskip

\textbf{Table III:} The scaled depth $\epsilon$ of the effective attractive
well and the strength of the effective adhesive interaction $\tau$ at pH $4.5$
as a function of the ionic strength $I$ and volume fraction $\eta$. The values
of $\epsilon$ and $\tau$ have been evaluated from Eqs. (\ref{vale}) and
(\ref{valt})\newpage

\hspace{0.3in}%
\begin{tabular}
[c]{|l|l|l|l|l|l|l|}\hline
$\eta$ &  & $0.15\text{ M}$ & $0.2\text{ M}$ & $0.25\text{ M}$ & $0.3\text{
M}$ & $0.45\text{ M}$\\\hline
$0.05$ & $\frac{\Pi}{\rho k_{B}T}$ & $1.143$ & $1.033$ & $0.949$ & $0.889$ &
$0.763$\\\hline
& $-2\Delta$ & $0.004$ & $0.001$ & $0.0004$ & $0.0001$ & $0.000008$\\\hline
$0.1$ & $\frac{\Pi}{\rho k_{B}T}$ & $1.290$ & $1.074$ & $0.915$ & $0.805$ &
$0.575$\\\hline
& $-2\Delta$ & $0.019$ & $0.006$ & $0.002$ & $0.0005$ & $0.00002$\\\hline
$0.15$ & $\frac{\Pi}{\rho k_{B}T}$ & $1.437$ & $1.123$ & $0.898$ & $0.749$ &
$0.448$\\\hline
& $-2\Delta$ & $0.044$ & $0.014$ & $0.004$ & $0.001$ & $0.00002$\\\hline
$0.2$ & $\frac{\Pi}{\rho k_{B}T}$ & $1.583$ & $1.183$ & $0.904$ & $0.721$ &
$0.375$\\\hline
& $-2\Delta$ & $0.085$ & $0.026$ & $0.008$ & $0.003$ & $0.000007$\\\hline
$0.3$ & $\frac{\Pi}{\rho k_{B}T}$ & $1.866$ & $1.361$ & $0.988$ & $0.753$ &
$0.340$\\\hline
& $-2\Delta$ & $0.228$ & $0.068$ & $0.022$ & $0.008$ & $0.00002$\\\hline
$0.4$ & $\frac{\Pi}{\rho k_{B}T}$ & $2.17$ & $1.659$ & $1.231$ & $0.960$ &
$0.470$\\\hline
& $-2\Delta$ & $0.488$ & $0.143$ & $0.046$ & $0.016$ & $0.0001$\\\hline
\end{tabular}
\smallskip

\textbf{Table IV:} The osmotic pressure from Eq. (\ref{osmpr})\ and its
correction from Eq. (\ref{crr}) as a function of the ionic strength $I$ and
the packing fraction $\eta$.\newpage

\section*{Figure Captions}

\textbf{Fig. 1:} The integrand of Eq.\ (\ref{vir}) versus the distance $r$. As
shown by the shaded regions, the repulsive tail is compensated by part of the
attractive interaction.

\textbf{Fig. 2:} Experimental data of the second virial coefficient $B_{2}$ of
lysozyme as a function of the ionic strength $I$ at a pH of about 4.5. The
second virial coefficient is scaled by the hard sphere value $B_{2}^{HS}%
$.\ Black squares: Bonnet\'{e} et al. \cite{BON}, pH 4.5, 20$%
\operatorname{{}^{\circ}{\rm C}}%
$; grey triangles: Curtis et al. \cite{CUR2}, pH 4.5, 20$%
\operatorname{{}^{\circ}{\rm C}}%
$; grey squares: Muschol et al. \cite{MUS}, pH 4.7, 20$%
\operatorname{{}^{\circ}{\rm C}}%
$; black stars: Curtis et al. \cite{CUR1}, pH 4.5, 25$%
\operatorname{{}^{\circ}{\rm C}}%
$; black diamonds: Bonnet\'{e} et al. \cite{BON}, pH 4.5, 25$%
\operatorname{{}^{\circ}{\rm C}}%
$; black triangles: Velev et al. \cite{VEL}, pH 4.5, 25$%
\operatorname{{}^{\circ}{\rm C}}%
$; white squares: Rosenbaum et al. \cite{ROS1}, pH 4.6, 25$%
\operatorname{{}^{\circ}{\rm C}}%
$; white diamonds: Rosenbaum et al. \cite{ROS2}, pH 4.6, 25$%
\operatorname{{}^{\circ}{\rm C}}%
$; grey stars: Bloustine et al. \cite{BLO}, pH 4.6, 25$%
\operatorname{{}^{\circ}{\rm C}}%
$; white stars: Piazza et al. \cite{PIA1}, pH 4.7, 25$%
\operatorname{{}^{\circ}{\rm C}}%
$; white triangles: Behlke et al. \cite{BEH}, pH 4.5; grey diamonds: Bloustine
et al. \cite{BLO}, pH 4.7. In all cases, the electrolyte is NaCl, often with a
small amount of Na acetate added.

\textbf{Fig. 3:} A fit of Eq. \ref{vir2} to the experimental data of Fig. 2
(except for those of Refs. \cite{PIA1} and \cite{BEH}). On the right-hand side
of the figure, the upper solid line corresponds to $I_{\theta}=0.19$,
$\delta=0.564$ and $U_{A}=1.48$, the upper dotted line to $I_{\theta}=0.20$,
$\delta=0.468$ and $U_{A}=1.70$ and the middle solid line to $I_{\theta}%
=0.21$, $\delta=0.379$ and $U_{A}=1.95$, all at an effective charge $Z_{eff}$.
The middle dotted line corresponds to $I_{\theta}=0.19$, $\delta=0.25$ and
$U_{A}=2.4$, the lower solid one to $I_{\theta}=0.20$, $\delta=0.167$ and
$U_{A}=2.87$ and the lower dotted one to $I_{\theta}=0.21$, $\delta=0.079$ and
$U_{A}=3.70$, all at a lowered effective charge $\overline{Z}$.

\textbf{Fig. 4:} Experimental data of the second virial coefficient $B_{2}$ of
lysozyme as a function of the ionic strength $I$ at a pH of about 7.5. The
second virial coefficient is scaled by the hard sphere value $B_{2}^{HS}$.
Black stars: Rosenbaum et al. \cite{ROS1}, pH 7.4, 25$%
\operatorname{{}^{\circ}{\rm C}}%
$; black triangles: Velev et al. \cite{VEL}, pH 7.5, 25$%
\operatorname{{}^{\circ}{\rm C}}%
$; black squares: Rosenbaum et al. \cite{ROS2}, pH 7.8, 25$%
\operatorname{{}^{\circ}{\rm C}}%
$.

\textbf{Fig. 5:} Fits of Eq. (\ref{benb2}) to experimental data of Fig. 3.
Full line ($Z_{eff}$ and $\delta\exp U_{A}=4.2$); Dotted line ($\overline{Z}$
and $\delta\exp U_{A}=3.7$).

\textbf{Fig. 6:} Comparison between the experimental data at pH 7.5 and full
theory Eq. (\ref{vir2}). Parameters as in the lower dotted curve in Fig. 3
($\delta=0.079$ and $U_{A}=3.70$).

\textbf{Fig. 7:} Ionic-strength dependence of AHS parameter $\tau$ at pH 4.5
and pH 7.5. The dotted line denotes the limiting value of $\tau$ as
$I\rightarrow\infty$.

\textbf{Fig. 8:} Inverse osmotic compressibility as a function of the volume
fraction $\eta$ at various ionic strengths. Experimental data: black squares:
$I=0.18$ M; black triangles: $I=0.23$ M; black stars: $I=0.28$ M; black
diamonds: $I=0.33$ M; open squares: $I=0.48$ M. All data from Rosenbaum et al.
\cite{ROS2}, except for those at $I=0.23$ M (black triangles) (Piazza et al.
\cite{PIA2}). Curves computed from Eq. (\ref{compr}) with $\delta=0.079$,
$U_{A}=3.70$ and the lowered effective charge $\overline{Z}$; $\tau$ has been
determined from Eq. (\ref{valt}). From top to bottom: $I=0.18$ M, $I=0.23$ M,
$I=0.28$ M, $I=0.33$ M and $I=0.48$ M.

\textbf{Fig. 9:} Dependence of $\ln\xi$ on $\mu$ at pH 4.5 and pH
7.5. In both cases $d\ln\xi/d\mu<0$ and
$d^{2}\ln\xi/d\mu^{2}\gtrsim0$ if $1\leq\mu\leq8$.\newpage

\begin{figure}[htbp]
\centering \epsfig{file=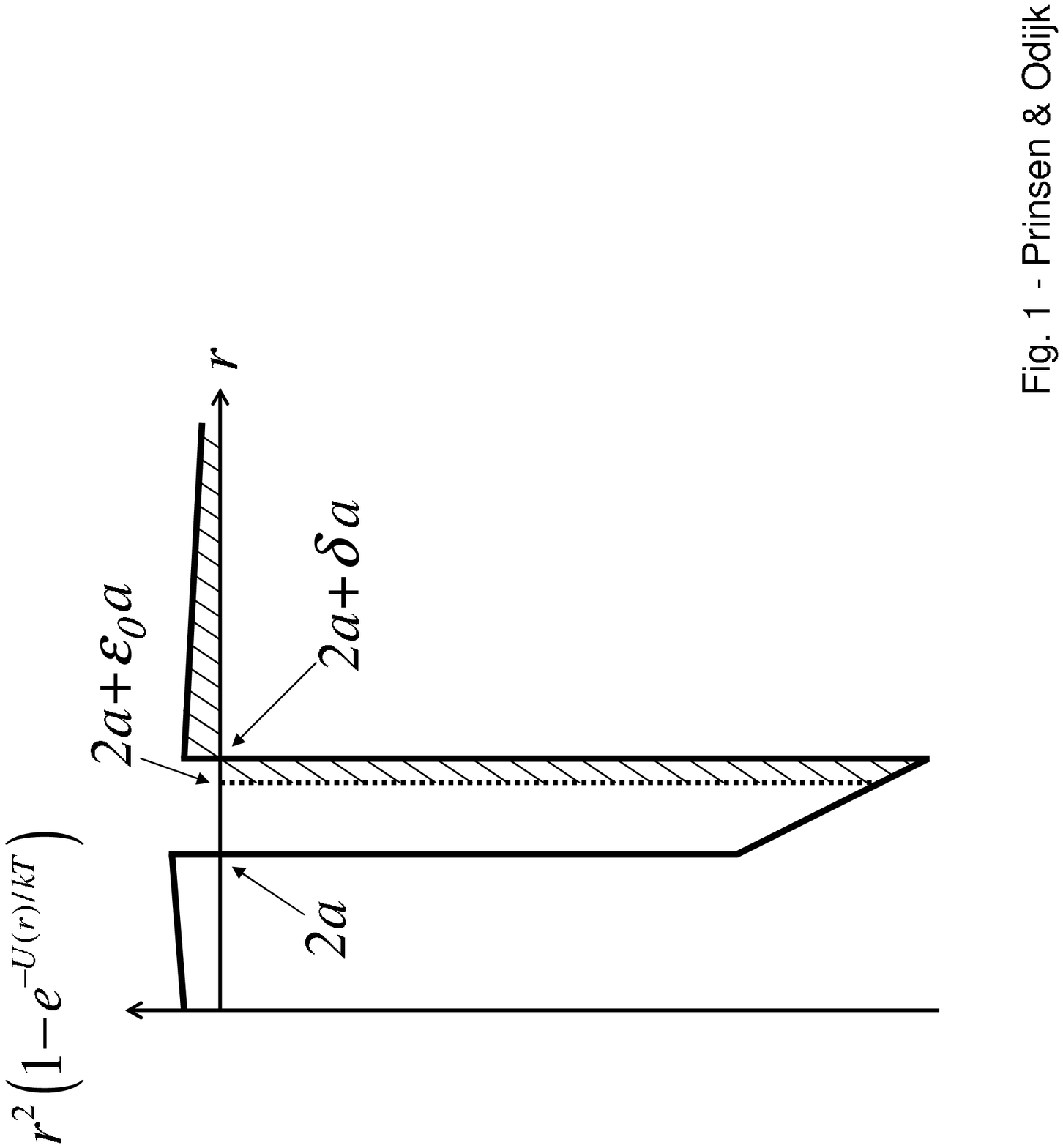, bbllx=70, bblly=180, bburx=580,
bbury=740}
\end{figure}\newpage

\begin{figure}[htbp]
\centering \epsfig{file=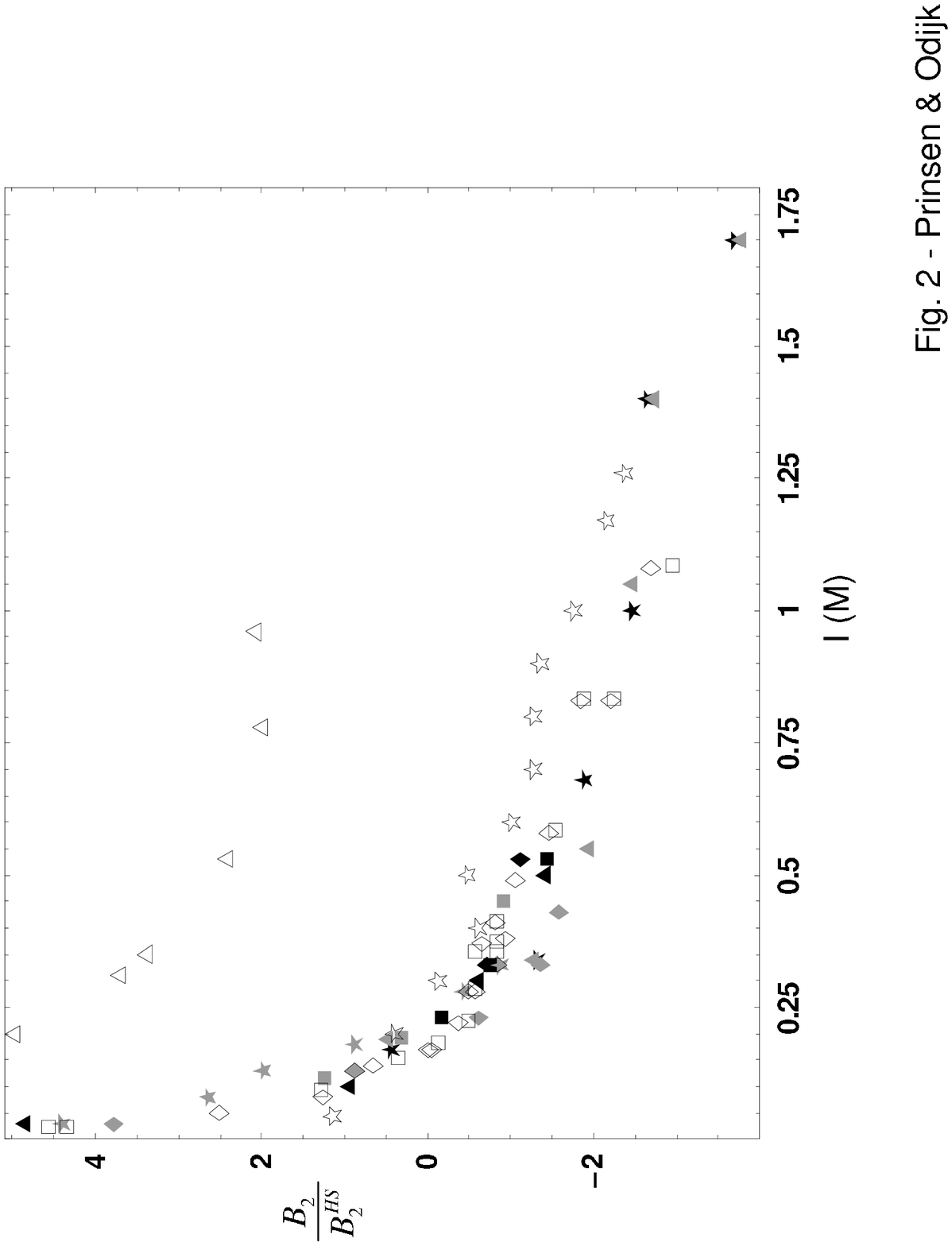, bbllx=60, bblly=60, bburx=580,
bbury=740}
\end{figure}\newpage

\begin{figure}[htbp]
\centering \epsfig{file=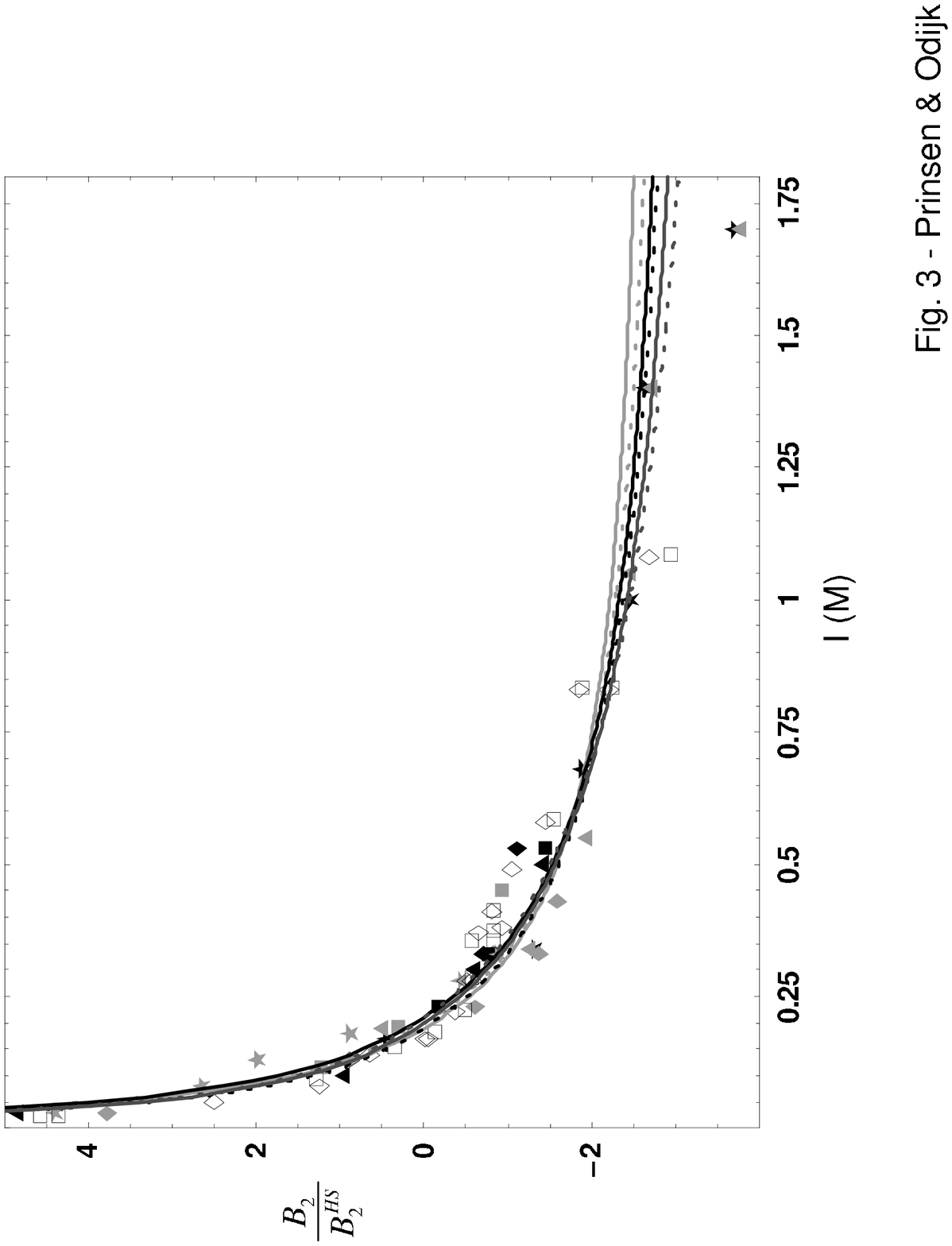, bbllx=60, bblly=60, bburx=580,
bbury=740}
\end{figure}\newpage

\begin{figure}[htbp]
\centering \epsfig{file=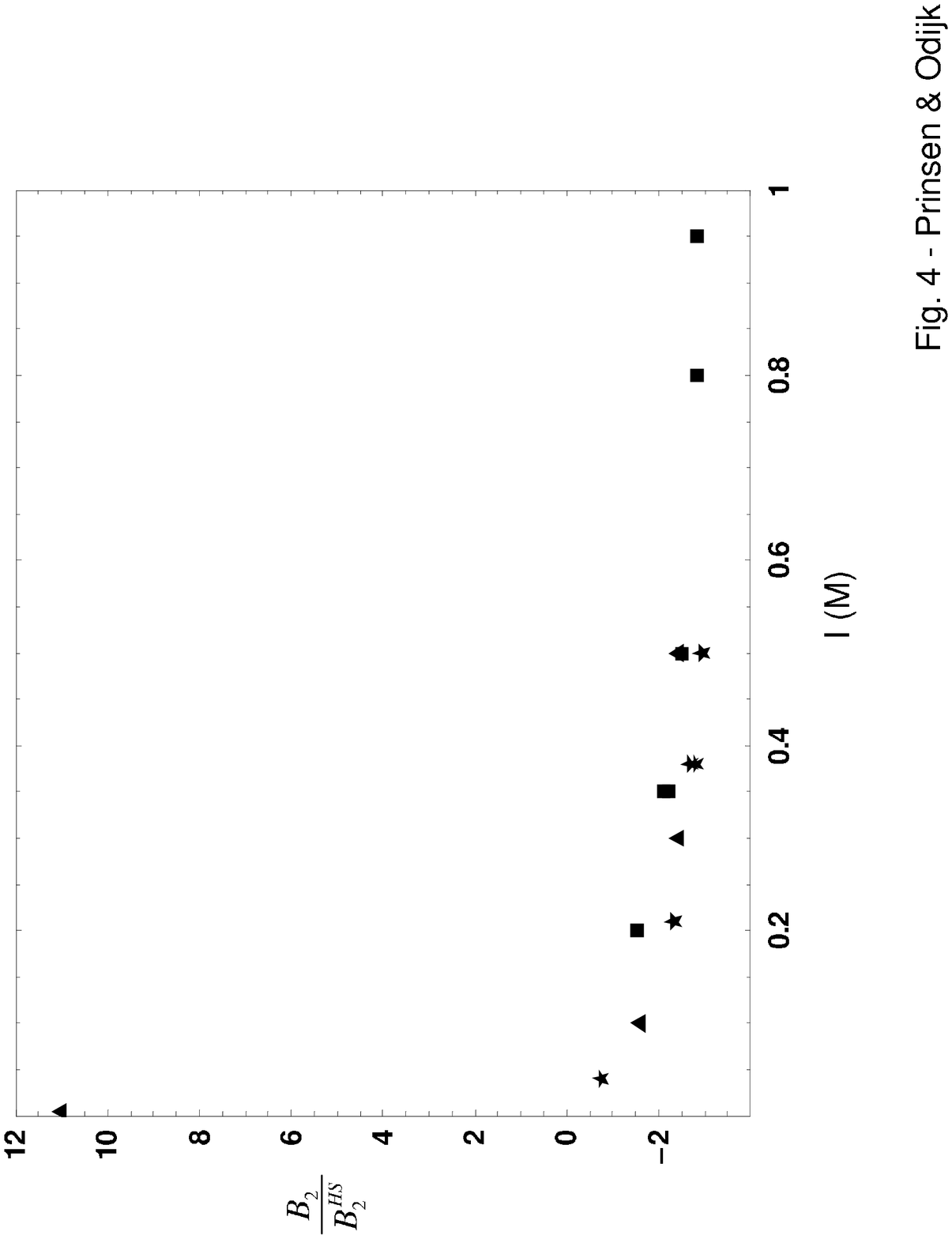, bbllx=60, bblly=60, bburx=580,
bbury=740}
\end{figure}\newpage

\begin{figure}[htbp]
\centering \epsfig{file=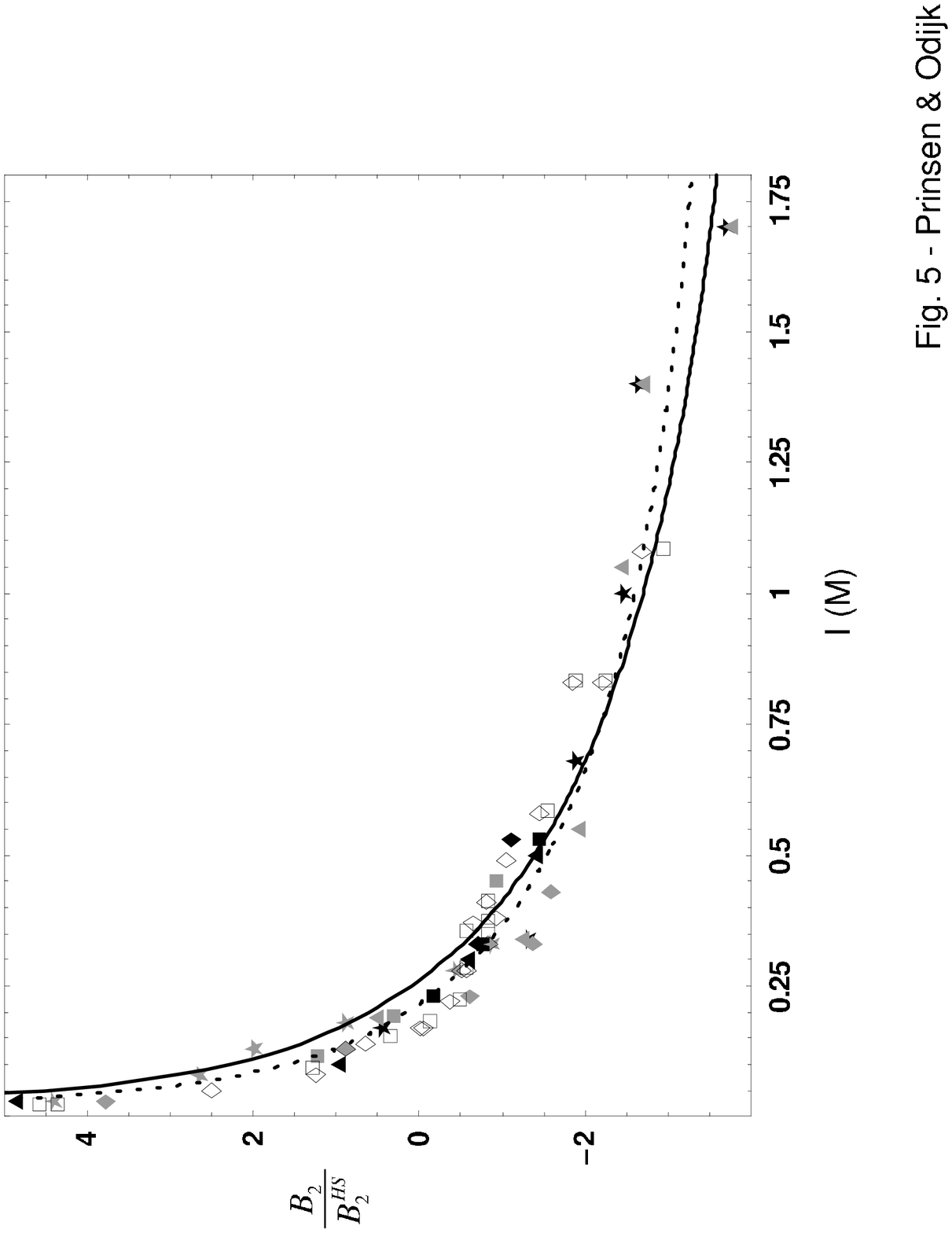, bbllx=60, bblly=60, bburx=580,
bbury=740}
\end{figure}\newpage

\begin{figure}[htbp]
\centering \epsfig{file=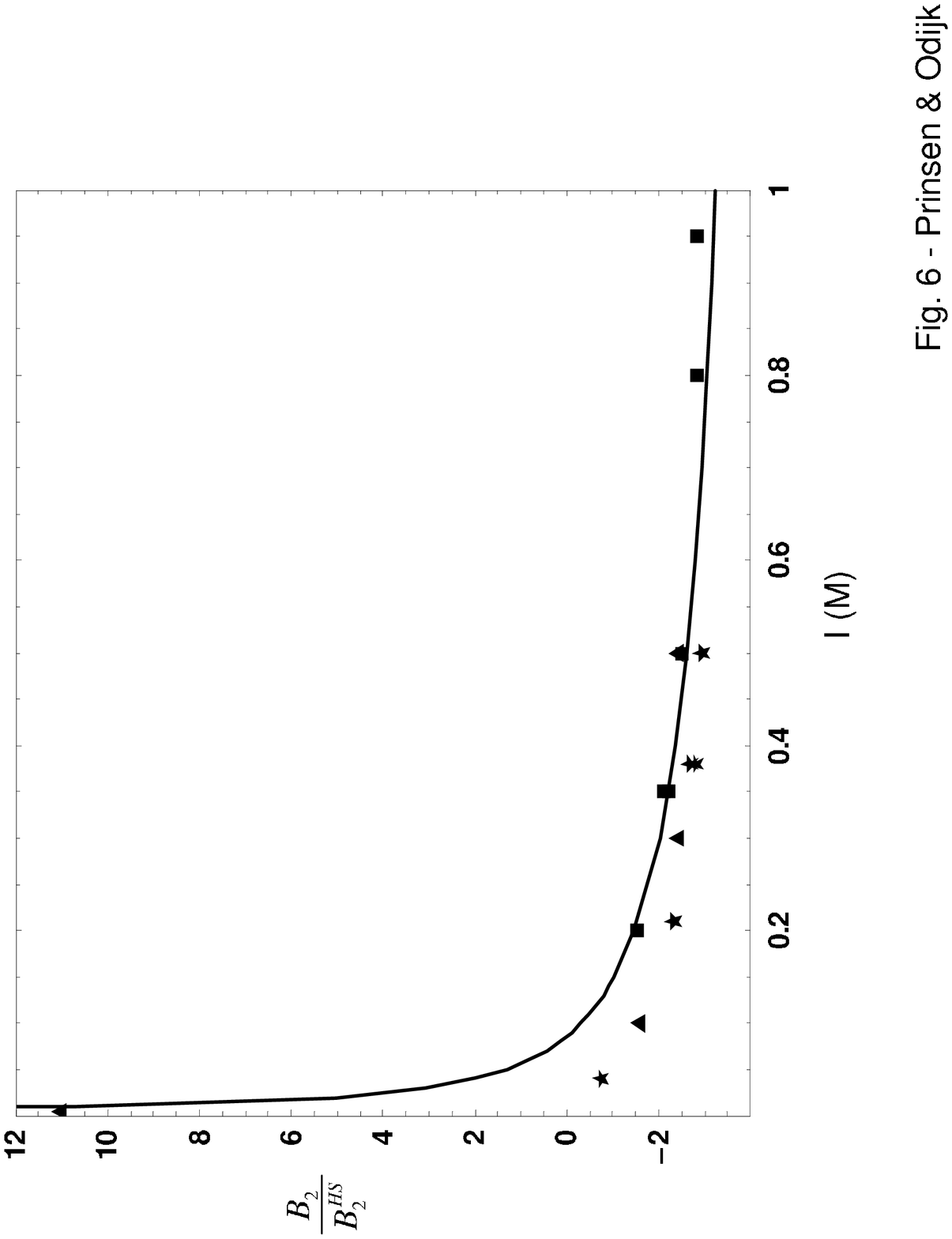, bbllx=60, bblly=60, bburx=580,
bbury=740}
\end{figure}\newpage

\begin{figure}[htbp]
\centering \epsfig{file=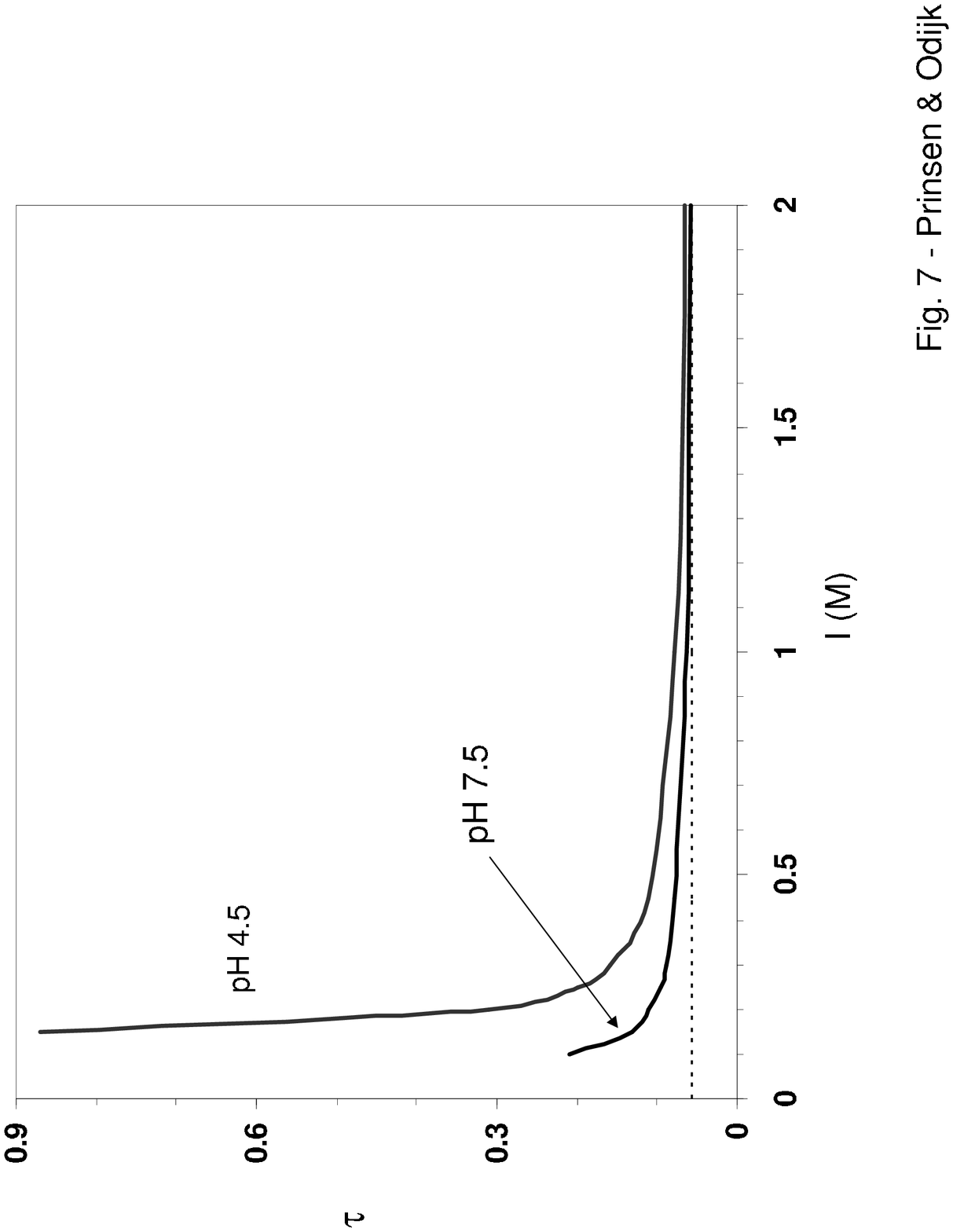, bbllx=60, bblly=60, bburx=580,
bbury=740}
\end{figure}\newpage

\begin{figure}[htbp]
\centering \epsfig{file=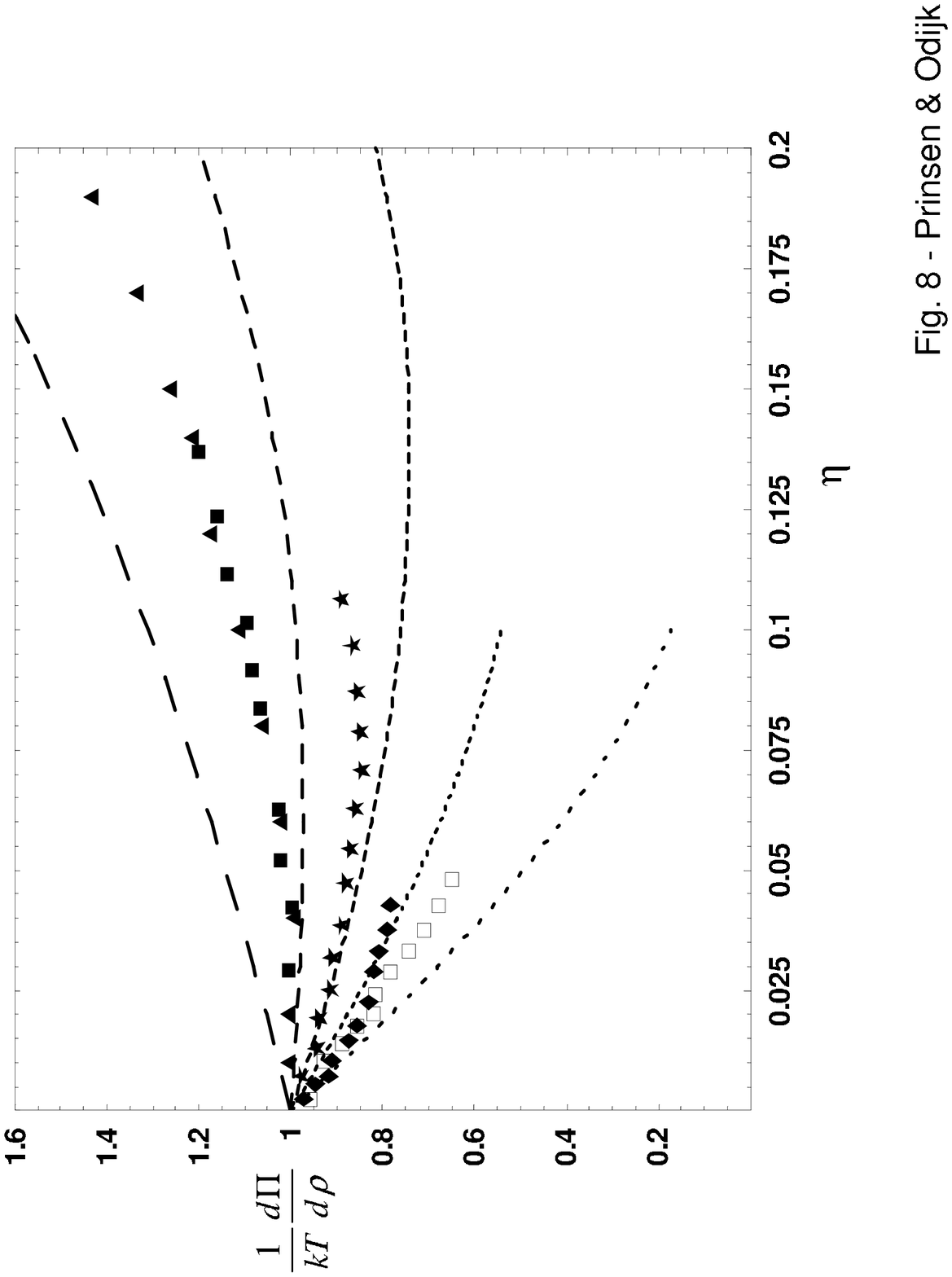, bbllx=60, bblly=50, bburx=580,
bbury=740}
\end{figure}\newpage

\begin{figure}[htbp]
\centering \epsfig{file=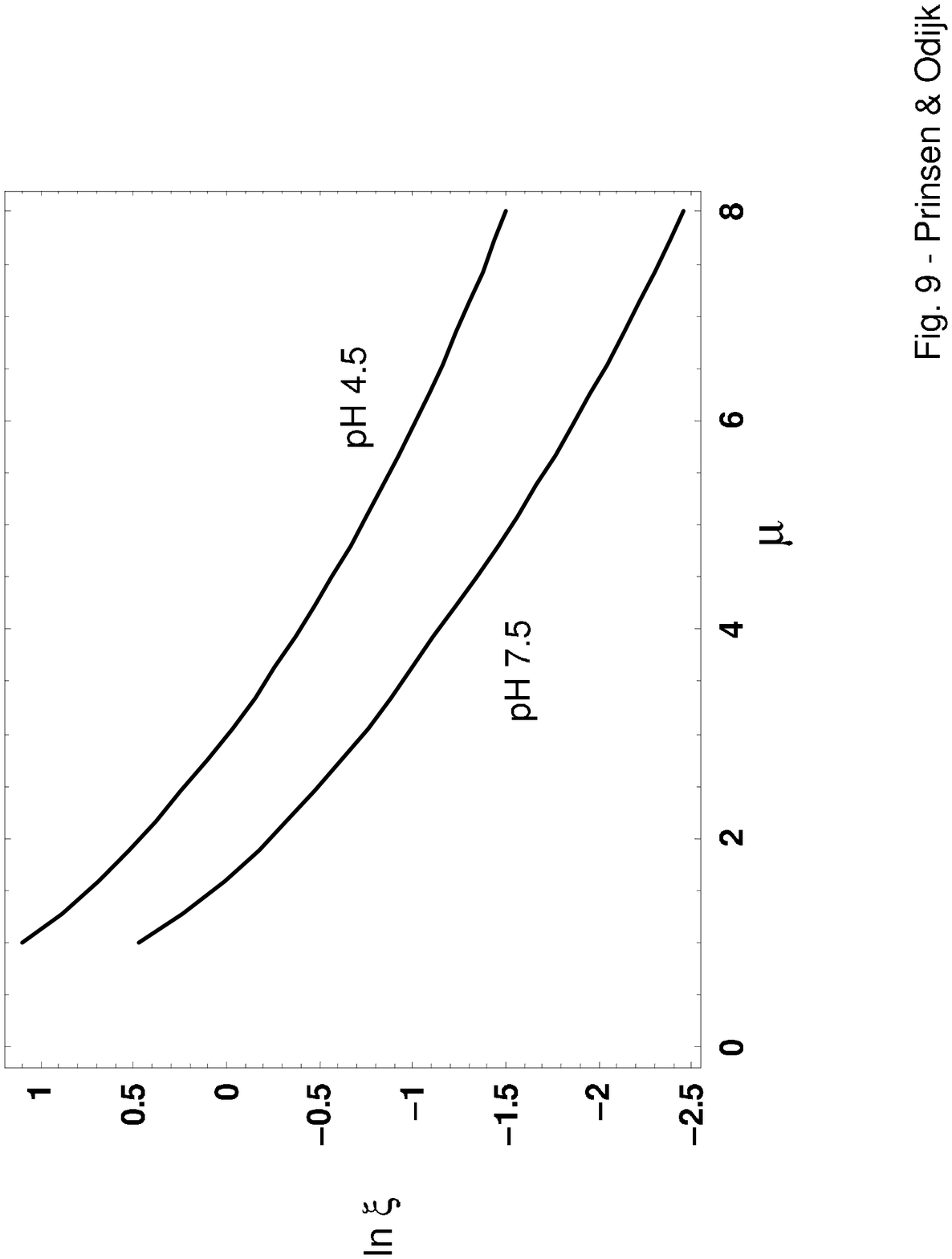, bbllx=60, bblly=60, bburx=580,
bbury=740}
\end{figure}\newpage

\end{document}